\documentclass{LMCS}

\usepackage{latexsym,psfrag,graphics,color,exscale,amssymb}
\usepackage[arrow,curve,frame,matrix,ps,dvips,pdftex]{xy}

\newcommand{\Nset}{\mathbb{N}}
\newcommand{\Nseto}{\Nset_0}

\newcommand{\A}{\mathcal{A}}
\newcommand{\B}{\mathcal{B}}

\newcommand{\T}{\mathcal{T}}
\newcommand{\G}{\mathcal{G}}
\newcommand{\F}{\mathcal{F}}

\renewcommand{\H}{\mathcal{H}}

\newcommand{\M}{\mathcal{M}}

\newcommand{\I}{\mathcal{I}}
\newcommand{\X}{\mathcal{X}}
\newcommand{\U}{\,\mathcal{U}\,}
\newcommand{\V}{\,\mathcal{V}\,}
\newcommand{\calP}{\mathcal{P}}

\newcommand{\tran}[1]{\stackrel{#1}{\rightarrow}}
\newcommand{\ltran}[1]{\stackrel{#1}{\longrightarrow}}

\newcommand{\St}{\mathit{St}}
\newcommand{\Stack}{\mathit{Stack}}
\newcommand{\chop}{\mathit{Chop}}
\newcommand{\Prob}{\mathit{Prob}}

\newcommand{\kw}[1]{\textbf{#1}}
\newcommand{\rv}[2]{V^{({#1})}_{#2}}
\newcommand{\MH}[1]{M_{#1}} 
\newcommand{\restrict}[2]{#1|_{#2}}

\newcommand{\fpath}{\mathit{FPath}}
\newcommand{\run}{\mathit{Run}}
\newcommand{\irun}{\mathit{IRun}}
\newcommand{\conf}{\mathcal{C}}
\newcommand{\Acc}{\mathit{Acc}}
\newcommand{\sat}{\mathit{Sat}}
\newcommand{\Var}[1]{{\langle #1 \rangle}}
\newcommand{\Pro}[1]{{[ #1 ]}}
\newcommand{\AP}{\mathit{Ap}}
\newcommand{\Cl}{\mathit{Cl}}
\newcommand{\Gen}{\mathit{Gen}}
\newcommand{\Obs}{\mathit{Obs}}
\newcommand{\States}{\mathit{States}}
\newcommand{\cosem}{\mbox{$[\![$}}
\newcommand{\cfsem}{\mbox{$]\!]$}}

\newcommand{\mylabel}{a}
\newcount\cvycet
\renewenvironment{itemize}%
  {\advance\cvycet by 1
   \ifnum\cvycet=1 \renewcommand{\mylabel}{$\bullet$}\fi%
   \ifnum\cvycet=2 \renewcommand{\mylabel}{$-$}\fi%
   \ifnum\cvycet=3 \renewcommand{\mylabel}{$*$}\fi%
   \begin{list}{\mylabel}{\setlength{\itemsep}{0pt}\setlength{\topsep}{1ex}%
        \setlength{\labelwidth}{2ex}\setlength{\leftmargin}{2ex}%
        \setlength{\labelsep}{.5ex}\setlength{\itemindent}{0ex}%
        \setlength{\parsep}{0pt}\setlength{\parskip}{0pt}%
        \setlength{\partopsep}{0pt}\setlength{\listparindent}{0pt}}}%
  {\end{list}\advance\cvycet by -1}

\def\doi{2 (1:2) 2006}
\lmcsheading%
{\doi}
{31}
{}
{}
{Dec.~\phantom{0}9, 2004}
{Mar.~\phantom{0}6, 2006}
{}

\begin{document}

\title{Model Checking Probabilistic Pushdown Automata}

\author[J.~Esparza]{Javier Esparza\rsuper a}
\address{{\lsuper a}Institute for Formal Methods in Computer 
Science, University of Stuttgart,
Universit\"{a}tsstr.~38, 70569 Stuttgart, Germany.}
\email{esparza@informatik.uni-stuttgart.de}
\thanks{{\lsuper a}Partially supported by the DFG-project ``Algorithms for
  Software-Model-Checking'' and by the EPSRC-Grant GR/93346 ``An
  Automata-theoretic Approach to Software-Model-Checking''.} 

\author[A.~Ku\v{c}era]{Anton\'{\i}n Ku\v{c}era\rsuper b}
\address{{\lsuper b}Faculty of Informatics, Masaryk 
  University, Botanick\'a 68a, CZ-60200 Brno, Czech Republic}
\email{tony@fi.muni.cz} 
\thanks{{\lsuper b}On leave at the Institute for Formal Methods in Computer Science, 
  University of Stuttgart. Supported by the Alexander von Humboldt 
  Foundation and by the research center Institute for 
  Theoretical Computer Science (ITI), project No.~1M0021620808.}

\author[R.~Mayr]{Richard Mayr\rsuper c}
\address{{\lsuper c}Department of Computer Science,
  North Carolina State University,
  900 Main Campus Drive,
  Campus Box 8207,
  Raleigh NC 27695, USA.}
\email{mayr@csc.ncsu.edu} 
\thanks{{\lsuper c}Supported by Landesstiftung
     Baden--W\"urttemberg, grant No.~21--655.023.}

\keywords{Pushdown automata, Markov chains, probabilistic model checking}
\subjclass{D.2.4, F.1.1, G.3}

\begin{abstract}
\noindent
We consider the model checking problem for probabilistic pushdown
automata (pPDA) and properties expressible in various probabilistic logics.
We start with properties that can be formulated as instances of a
generalized random walk problem. We prove that 
both qualitative and quantitative model checking for this class of 
properties and pPDA is decidable. Then we show that model checking for 
the qualitative fragment of the logic PCTL and pPDA is also decidable.
Moreover, we develop 
an error-tolerant model checking algorithm for PCTL 
and the subclass of stateless pPDA. Finally, we consider the class
of $\omega$-regular properties and show that both qualitative and 
quantitative model checking for pPDA is decidable.
\end{abstract}

\maketitle
\section{Introduction}\label{sec:introduction}

Probabilistic systems can be used for modeling systems that exhibit
uncertainty, such as communication protocols over unreliable channels,
randomized distributed systems, or fault-tolerant systems.
Finite-state models of such systems often use variants of
probabilistic automata whose underlying semantics is defined in terms
of homogeneous Markov chains, which are also called ``fully
probabilistic transition systems'' in this context.  For fully probabilistic
finite-state systems, algorithms for various (probabilistic) temporal
logics like LTL, PCTL, PCTL$^*$, probabilistic $\mu$-calculus, etc.,
have been presented
in \cite{LS:time-chance-IC,HS:Prob-temp-logic,Vardi:verif-prob-conc-syst,%
CY:temporal-probabilistic-FOCS,HJ:logic-time-probability-FAC,%
ASBBV:logic-stochastic-CAV,CY:probab-verification-JACM,%
HK:quantitative-mu-calculus-LICS,CSS:Prob-LTL-checking}. 
As for infinite-state systems, most
works so far considered probabilistic lossy channel systems 
\cite{IN:probLCS-TACAS} which
model asynchronous communication through unreliable channels
\cite{BE:probLCS-algorithms-ARTS,ABIJ:CONCUR00-IC,%
AR:prob-systems-faulty-FOSSACS,BS:PLCS-prob-decidable}. 
A notable recent result is the
decidability of quantitative model checking of liveness properties
specified by B\"uchi-automata for probabilistic lossy channel systems
\cite{Rabinovich:probLCS-LTL-ICALP}. In fact, this algorithm is \emph{error
  tolerant} in the sense that the quantitative model checking is
solved only up to an arbitrarily small (but non-zero) given error.

In this paper we consider \emph{probabilistic pushdown automata (pPDA)},
which are a natural model for probabilistic sequential 
programs with possibly recursive procedure calls. There is a large number of
results about model checking of non-probabilistic PDA or
similar models (see for instance
\cite{AEY:recursive-state-machines,BS:mu-calculus,%
EHRS:MC-PDA,Walukiewicz:CAV96-IC}), 
but the probabilistic extension has so far not been considered.
As a related work we can mention \cite{MO:probPDA-properties-TCS}, 
where it is shown that a restricted subclass of pPDA
(where essentially all probabilities for outgoing arcs are either $1$ or 
$1/2$) generates a richer class of languages than non-deterministic PDA.
Another work \cite{AAP:prob-grammars-automata-ACL} shows
the equivalence of pPDA and probabilistic context-free grammars.  
There are also recent results of 
\cite{BKS:pPDA-temporal,EY:RMC-SG-equations,EY:pPDA-LTL-complexity} 
which are directly related to the results presented in this paper. 
A detailed discussion is postponed to Section~\ref{sec-concl}.

Here we consider model checking problems for pPDA and its natural subclass
of \emph{stateless pPDA} denoted pBPA\footnote{This is a standard 
notation adopted in 
concurrency theory. The subclass of stateless PDA corresponds to a 
natural subclass of ACP known as Basic Process Algebra \cite{BW:book}.} 
and various probabilistic logics.
\begin{figure}[http]
\[\xy;<22 pt,0 pt>:
  (0,1)*+{\ldots}="A",
  (2,1)*{\cir<8 pt>{}}="B",
  (4,1)*{\cir<8 pt>{}}="C",
  (6,1)*{\cir<8 pt>{}}="D",
  (8,1)*+{\ldots}="E",
  (0,0)*{\scriptstyle DDZ},
  (2,0)*{\scriptstyle DZ},
  (4,0)*{\scriptstyle Z},
  (6,0)*{\scriptstyle IZ},
  (8,0)*{\scriptstyle IIZ},
  \ar@<3 pt>^x"A";"B"
  \ar@<3 pt>^x"B";"C"
  \ar@<3 pt>^x"C";"D"
  \ar@<3 pt>^x"D";"E"
  \ar@<3 pt>^{1-x}"E";"D"
  \ar@<3 pt>^{1-x}"D";"C"
  \ar@<3 pt>^{1-x}"C";"B"
  \ar@<3 pt>^{1-x}"B";"A"
\endxy
\]
\caption{Bernoulli random walk as a pBPA}
\label{fig-walk}
\end{figure}
We start with a class of properties that can be specified as 
a generalized \emph{random walk problem}. To get a better intuition about 
this class of problems, realize that some  random walks
can easily be specified by pBPA
systems. For example, consider a pBPA with just three {\em stack symbols}
$Z,I,D$ and {\em transitions} $Z \tran{x} IZ$, $Z \tran{1-x} DZ$, 
$I \tran {x} II$,
$I \tran{1-x} \varepsilon$, $D \tran{1-x} DD$, and $D \tran{x} \varepsilon$,
where $x \in [0,1]$ and $\varepsilon$ denotes the empty string. A
transition $X \tran{x} w$ means that if the current top stack symbol is $X$,then it can be replaced by $w$ with probability $x$. The transition 
graph of this pBPA with $Z$ as initial stack content 
(see Fig.~\ref{fig-walk}) is the well-known \emph{Bernoulli walk}. 
A typical question examined in the theory of random walks is  
``Do we eventually revisit a given state (with probability one)?'', or more 
generally ``What is the probability of reaching a given state from 
another given state?'' For example, it is a standard result that the
state $Z$ of Fig.~\ref{fig-walk} is revisited with probability $1$ if{}f
$x = 1/2$. This simple example indicates that answers to qualitative 
questions about pPDA (i.e., whether something holds with probability 
$1$ or $0$) depend on the exact probabilities of individual transitions.
This is different from finite-state systems where qualitative properties
depend only on the topology of a given finite-state Markov chain
\cite{HJ:logic-time-probability-FAC}. 

The generalized 
random walk problem is formulated as follows: Let $\conf_1$ and $\conf_2$
be subsets of the set of states of a given Markov chain, and let $s$
be a state of $\conf_1$. What is the probability that
a run initiated in $s$ hits a state of $\conf_2$ via a path leading only 
through the states of $\conf_1$?
Let us denote this probability by $\calP(s,\conf_1 \U \conf_2)$.
The problem of computing $\calP(s,\conf_1 \U \conf_2)$ 
has previously been considered (and solved) for finite-state
systems, where this probability can be computed precisely 
\cite{HJ:logic-time-probability-FAC,CY:probab-verification-JACM}.
In Section~\ref{sec-random}, we propose a solution for pPDA
applicable to those sets $\conf_1,\conf_2$ which are \emph{regular}, i.e.,
recognizable by finite-state automata (realize that pPDA configurations
can be written as words of the form $p\alpha$, where $p$ is a control state
and $\alpha$ a sequence of stack symbols). More precisely, we show that
the problem whether $\calP(s,\conf_1 \U \conf_2) \sim \varrho$, where
${\sim} \in \{{\leq},{<},{\geq},{>},{=}\}$ and 
$\varrho \in [0,1]$, is decidable.
Interestingly, this is achieved without explicitly computing 
the probability $\calP(s,\conf_1 \U \conf_2)$. Moreover,
for an arbitrary precision $0 < \lambda < 1$ we can compute rational
lower and upper approximations $\calP^\ell,\calP^u \in [0,1]$ such that 
$\calP^\ell \leq \calP(s,\conf_1 \U \conf_2) \leq \calP^u$ and 
$\calP^u - \calP^\ell \leq \lambda$. 

In Section~\ref{sec-PCTL}, we consider the model checking problem for
pPDA and the logic PCTL. This is a more general problem
than the one about random walks (the class of properties
expressible in PCTL is strictly larger). 
In Section~\ref{sec-PCTL-quality}, we give a model checking algorithm
for the \emph{qualitative fragment} of PCTL and pPDA processes.
For general PCTL formulas and pBPA processes, an \emph{error tolerant} 
model checking algorithm is developed in Section~\ref{sec-PCTL-nBPA}.
The question whether this result can be extended to pPDA is left open.

Finally, in Section~\ref{sec-Buchi} we prove that both qualitative and
quantitative model checking for the class of $\omega$-regular
properties is decidable for pPDA. In \cite{EKM:prob-PDA-PCTL}, it was
shown that the qualitative and quantitative model-checking problem is
decidable for pPDA and a subclass of $\omega$-regular properties that
are definable by deterministic B\"{u}chi automata. Later, it has been
observed in \cite{BKS:pPDA-temporal} that the technique can easily be
generalized to Muller automata, and thus the decidability result was
extended to all $\omega$-regular properties 
(in \cite{BKS:pPDA-temporal}, some complexity results were also
presented).  The construction presented in this paper is a slightly 
generalized and polished version of the algorithms given in
\cite{EKM:prob-PDA-PCTL,BKS:pPDA-temporal}, which can now be seen as
instances of a more abstract result.

In Section~\ref{sec-concl} we conclude by remarks on open problems and
recent related work of
\cite{BKS:pPDA-temporal,EY:RMC-SG-equations,EY:pPDA-LTL-complexity}.

\section{Preliminary Definitions}
\label{sec-defs}
\begin{defi} A  \emph{probabilistic transition system} 
  is a triple $\T = (S,\tran{},\Prob)$ where $S$ is a finite or
  countably infinite set of \emph{states}, ${\tran{}} \subseteq S
  \times S$ is a \emph{transition relation}, and $\Prob$ is a function
  which to each transition $s \tran{} t$ of $\T$ assigns its
  probability $\Prob(s \tran{} t) \in (0,1]$ so that for every $s \in
  S$ we have \[\sum_{s \tran{} t} \Prob(s \tran{} t) \in \{0,1\} \]
  The sum above is $0$ if{}f $s$ does not have
  any outgoing transitions.
\end{defi}
In the rest of this paper we also write $s \tran{x} t$ instead of
$\Prob(s \tran{} t) = x$. A \emph{path} in $\T$ is a finite or
infinite sequence $w = s_0;s_1;\cdots$ of states such that $s_i
\tran{} s_{i+1}$ for every $i$. We also use $w(i)$ to denote the state
$s_i$ of $w$ (by writing $w(i) = s$ we implicitly impose the condition
that the length of $w$ is at least $i+1$).  A \emph{run} is a maximal
path, i.e., a path which cannot be prolonged. The sets of all finite
paths, all runs, and all infinite runs of $\T$ are denoted $\fpath$,
$\run$, and $\irun$, respectively\footnote{In this paper, $\T$ is
  always clear from the context.}.  Similarly, the sets of all finite
paths, runs, and infinite runs that start in a given $s \in S$ are
denoted $\fpath(s)$, $\run(s)$, and $\irun(s)$, respectively.

Each $w \in \fpath$ determines a \emph{basic cylinder} $\run(w)$ which
consists of all runs that start with $w$.  To every $s \in S$ we
associate the probabilistic space $(\run(s),\F,\calP)$ where $\F$ is
the $\sigma$-field generated by all basic cylinders $\run(w)$ such that
$w$ starts with $s$, and $\calP: \F \rightarrow [0,1]$ is the unique
probability function such that $\calP(\run(w)) = \Pi_{i=0}^{m-1} x_i$
where $w = s_0;\cdots;s_m$ and $s_i \tran{x_i} s_{i+1}$ for every $0
\leq i < m$ (if $m=0$, we put $\calP(\run(w)) = 1$).

\subsection{The Logic PCTL}

PCTL, the probabilistic extension of CTL, was defined 
in \cite{HJ:logic-time-probability-FAC}.  Let $\AP = \{a,b,c,\dots\}$
be a countably infinite set of \emph{atomic propositions}. The syntax
of PCTL\footnote{For simplicity we omit the bounded `until' operator of
  \cite{HJ:logic-time-probability-FAC}.} 
is given by the following abstract syntax equation:
\begin{eqnarray*}
  \varphi & ::= & \texttt{tt} \mid a \mid \neg \varphi \mid
      \varphi_1 \wedge \varphi_2 \mid
      \X^{\sim \varrho} \varphi \mid 
      \varphi_1 \U^{\sim \varrho} \varphi_2
\end{eqnarray*}
Here $a$ ranges over $\AP$, $\varrho \in [0,1]$, and 
${\sim} \in \{\leq,<,\geq,>\}$. Let $\T = (S,\tran{},\Prob)$
be a probabilistic transition system. For all $s \in S$,
all $\conf,\conf_1,\conf_2 \subseteq S$, and all $k \in \Nset_0$, let
\begin{itemize}
\item $\run(s,\X \conf) = \{w \in \run(s) \mid w(1) \in \conf\}$
\item $\run(s,\conf_1 \U \conf_2) = 
  \{w \in \run(s) \mid \exists i \geq 0 : w(i) \in \conf_2 \text{ and }
  w(j) \in \conf_1 \text{ for all } 0 \leq j <i\}$
\item $\fpath^k(s,\conf_1 \U \conf_2) = \{ s_0; {\cdots}; s_\ell \in \fpath(s) 
    \!\mid\! 0\leq \ell \leq k, s_\ell \in \conf_2 \text{ and } s_j \in 
    \conf_1 {\smallsetminus} \conf_2 \text{\ for all\ }  0 \leq j <\ell \}$
\item $\fpath(s,\conf_1 \U \conf_2) = 
   \bigcup_{k=0}^{\infty}\fpath^k(s,\conf_1 \U \conf_2)$
\end{itemize}
The set $\run(s,\X \conf)$ is clearly $\calP$-measurable, and the same
holds for $\run(s,\conf_1 \U \conf_2)$ because
$$
   \calP(\run(s,\conf_1 \U \conf_2)) = 
   \sum_{w \in \fpath(s,\conf_1 \U \conf_2)} \calP(\run(w)).
$$
In the rest of this paper, we will usually write $\calP(s,\X \conf)$ and
$\calP(s,\conf_1 \U \conf_2)$ instead of $\calP(\run(s,\X \conf))$
and $\calP(\run(s,\conf_1 \U \conf_2))$, respectively.

Let $\nu : \AP \rightarrow 2^S$ be a \emph{valuation}. The denotation
of a PCTL formula $\varphi$ over $\T$ w.r.t.\ $\nu$, denoted
$\cosem \varphi \cfsem^\nu$, is defined inductively
as follows:
\begin{eqnarray*}
  \cosem \texttt{tt} \cfsem^\nu & = & S\\
  \cosem a \cfsem^\nu & = & \nu(a)\\
  \cosem \neg \varphi \cfsem^\nu & = & 
   S \smallsetminus \cosem \varphi \cfsem^\nu\\
  \cosem \varphi_1 \wedge \varphi_2 \cfsem^\nu & = &  
    \cosem \varphi_1 \cfsem^\nu \cap \cosem \varphi_2 \cfsem^\nu\\
  \cosem \X^{\sim \varrho} \varphi \cfsem^\nu & = &
     \{ s \in S \mid \calP(s,\X \cosem \varphi \cfsem^\nu) 
     \sim \varrho\}\\
  \cosem \varphi_1 \U^{\sim \varrho} \varphi_2 \cfsem^\nu & = &
     \{ s \in S \mid \calP(s,\cosem \varphi_1\cfsem^\nu \U
          \cosem \varphi_2\cfsem^\nu) \sim \varrho\}
\end{eqnarray*}
As usual, we write $s \models^\nu \varphi$ instead of 
$s \in \cosem \varphi \cfsem^\nu$. 

The \emph{qualitative fragment} of PCTL
is obtained by restricting the allowed operator/ number combinations
to `$\leq 0$' and \mbox{`$\geq 1$'}, which will be also written as
`$=0$' and `$=1$', resp. (Observe that `$<1$', `$>0$' are definable
from `$\leq 0$', `$\geq 1$', and negation; for example, 
$a \U^{<1} b \equiv \neg (a \U^{\geq 1} b)$.) 

\subsection{Probabilistic PDA}

\begin{defi}
A \emph{probabilistic pushdown automaton (pPDA)} is a tuple 
$\Delta = (Q,\Gamma,\delta,\Prob)$
where $Q$ is a finite set of \emph{control states}, $\Gamma$ is a
finite \emph{stack alphabet}, 
$\delta \subseteq Q \times \Gamma \times Q \times \Gamma^*$ is 
a  finite \emph{transition relation} (we write $pX \tran{} q \alpha$ instead
of $(p,X,q,\alpha) \in \delta$), and $\Prob$ is a function which to 
each transition $pX \tran{} q\alpha$ assigns 
its probability $\Prob(pX \tran{} q\alpha) \in (0,1]$ and satisfies 
$\sum_{pX \tran{} q\alpha} \Prob(pX \tran{} q\alpha) \in \{0,1\}$
for all $p \in Q$ and $X \in \Gamma$.

A \emph{pBPA} is a pPDA with just one control state.
Formally, a pBPA is understood as a triple 
$\Delta = (\Gamma,\delta,\Prob)$ where 
$\delta \subseteq \Gamma \times \Gamma^*$. 
\end{defi}
In the rest of this paper we adopt a more intuitive notation, writing
$pX \tran{x} q\alpha$ instead of $\Prob(pX \tran{} q\alpha) = x$.
A \emph{configuration} of $\Delta$ is an element of $Q \times \Gamma^*$.
The set of all configurations of $\Delta$ is denoted
by $\conf(\Delta)$.
We also assume (w.l.o.g.) that if $pX \tran{} q\alpha \in \delta$, then
$|\alpha| \leq 2$. It is easy to transform an arbitrary pair
$(\Delta, F)$, where $\Delta$ is a pPDA and $F$ is a
a PCTL formula or $\omega$-property, into another pair $(\Delta', F')$
such that $\Delta'$ satisfies the assumption above and $\Delta$ satisfies
$F$ if and only if $\Delta'$ satisfies $F'$. Moreover, the transformation
takes linear time. For instance, a transition rule 
$pX \tran{x} qYZW$ of $\Delta$ is transformed
into two transitions $pX \tran{x} p'Y'W$ and $p'Y' \tran{1} qYZ$ in 
$\Delta'$, where $p', Y'$ are a fresh control state and a fresh 
stack symbol, respectively.

To $\Delta$ we associate the probabilistic transition system
$\T_\Delta$ where $\conf(\Delta)$ is the set of states and the probabilistic 
transition relation is determined as follows:
$pX \beta \tran{x} q \alpha \beta$ is a transition of $\T_\Delta$ 
iff $pX \tran{x} q\alpha$ is a transition of $\Delta$ and $\beta \in \Gamma^*$.

The model checking problem for pPDA configurations and PCTL formulate
(i.e., the question whether $p\alpha \models^\nu \varphi$ for given
$p\alpha$, $\varphi$, and $\nu$) is clearly undecidable for general
valuations. Therefore, we restrict ourselves to \emph{regular} valuations
which to every $a \in \AP$ assign a \emph{regular set of configurations}:

\begin{defi}
  A \emph{$\Delta$-automaton} is a triple $\A = (\St,\gamma,\Acc)$
  where $\St$ is a finite set of \emph{states} s.t. $Q \subseteq \St$, 
  $\gamma : \St \times \Gamma \rightarrow \St$ is a (total)
  \emph{transition function}, 
  and $\Acc \subseteq \St$ a set of \emph{accepting states}.

  The function $\gamma$ is extended to the elements of $\Gamma^*$ in
  the standard way. 
  Each $\Delta$-automaton $\A$ determines a set 
  $\conf(\A) \subseteq \conf(\Delta)$ given by $p\alpha \in \conf(\A)$
  if{}f $\gamma(p,\alpha^R) \in \Acc$. Here 
  $\alpha^R$ is the reverse of $\alpha$, i.e., the word obtained 
  by reading $\alpha$ from right to left.

  We say that a set $\conf \subseteq \conf(\Delta)$ is \emph{regular} 
  if{}f there is a $\Delta$-automaton $\A$ such that $\conf = \conf(\A)$.
\end{defi}

\noindent
In other words, regular sets of configurations are recognizable by
finite-state automata which read the stack bottom-up (the bottom-up
direction was chosen just for technical convenience).

An important technical step is that one can reduce the
model-checking problem for regular valuations to the problem
for {\em simple} valuations that assign to each atomic proposition
a {\em simple} set of configurations. Loosely speaking, a set of 
configurations is simple if we can decide whether a configuration
belongs to the set by inspecting only its control state and its top
stack symbol. 

\begin{defi}
\label{def-simple}
  A set of configurations $\conf \subseteq \conf(\Delta)$ is 
  \emph{simple} if there is
  a set $G \subseteq Q \times (\Gamma \cup \{\varepsilon\})$ such that
  for each $p\alpha \in \conf(\Delta)$ we have that 
  $p\alpha \in \conf$ if{}f either $\alpha = \varepsilon$ and 
  $p\varepsilon \in G$, or $\alpha = X\beta$ and $pX \in G$.
\end{defi}

The reason why we only need to consider simple valuations
is a bisimilarity property.
Let  $\conf_1,\cdots,\conf_k \subseteq \conf(\Delta)$ be regular sets 
of configurations, and assume that all we can observe from a 
configuration is whether it belongs to $\conf_i$ for every 
$1 \leq i \leq k$. Loosely
speaking, Lemma \ref{lem-reg-sim} below states that we can effectively
construct another pPDA $\Delta'$ and {\em simple} sets of configurations
$\conf_1',\cdots,\conf_k' \subseteq \conf(\Delta)$ such that $\Delta$ and $\Delta'$ are bisimilar with respect to these observables
(in the usual definition of bisimilarity one observes transitions between
configurations, while here we observe the configurations themselves,
but otherwise the notion is the same). The idea of the construction is to
take $\Delta$-automata $\A_1,\cdots,\A_k$ accepting the sets 
$\conf_1,\cdots,\conf_k$, and construct $\Delta'$ such that the
following holds: If the current configuration
of $\Delta$ is $p\alpha$, then in the simulating configuration
of $\Delta'$ the topmost stack symbol stores the states reached by 
the $\Delta$-automata after reading $\alpha^R$ from the initial state $p$.
Although this construction is standard 
(see, e.g., \cite{EKS:PDA-regular-valuations-IC}),
we include an explicit proof for the sake of completeness.

\begin{lem}
\label{lem-reg-sim}
  For each pPDA $\Delta = (Q,\Gamma,\delta,\Prob)$ and regular sets
  $\conf_1,\cdots,\conf_k \subseteq \conf(\Delta)$
  there effectively exists a pPDA  $\Delta' = (Q,\Gamma', \delta',\Prob')$, 
  simple sets $\conf'_1,\cdots,\conf'_k \subseteq \conf(\Delta')$, 
  and an injective mapping $\G : \conf(\Delta) \rightarrow \conf(\Delta')$ 
  such that for each $p\alpha \in \conf(\Delta)$ the following conditions
  are satisfied:
  \begin{itemize}
  \item for each $1 \leq j \leq k$ we have  $p\alpha \in \conf_j$ if{}f 
     $\G(p\alpha) \in \conf'_j$;
  \item if $p\alpha \tran{x} q\beta$, then $\G(p\alpha) \tran{x} \G(q\beta)$;
  \item if $\G(p\alpha) \tran{x} s$ for some $s \in \conf(\Delta')$, then
     there is $p\alpha \tran{x} q\beta$ such that $\G(q\beta) = s$.
  \end{itemize}
  Moreover, if $\conf \subseteq \conf(\Delta')$ is regular, then
  $\G^{-1}(\conf)$ is also regular.
\end{lem}
\begin{proof}
  For each $1 \leq i \leq k$, let $\A_i = (\St_i,\gamma_i,\Acc_i)$ be 
  a $\Delta$-automaton such that $\conf(\A_i) = \conf_i$. 
  Let $\States = \prod_{i=1}^k \prod_{p \in Q} \St_i$. For given
  $\vec{s} \in \States$, $1 \leq i \leq k$, and $p \in Q$, we 
  denote by $\vec{s}(i,p)$ the component of $\vec{s}$ which corresponds
  to $i$ and $p$.

  We put $\Gamma' = \Gamma \times \States$. The
  transition function $\delta'$ and probabilities $\Prob'$ are defined 
  as follows:
  \begin{itemize}
  \item if $pX \tran{x} q \varepsilon \in \delta$, then 
     $p (X,\vec{s}) \tran{x} q \varepsilon$ for each $\vec{s} \in \States$;
  \item if $pX \tran{x} qY \in \delta$, then
     $p (X,\vec{s}) \tran{x} q (Y,\vec{s})$ for each $\vec{s} \in \States$; 
  \item if $pX \tran{x} qYZ \in \delta$, then
     $p (X,\vec{s}) \tran{x} q (Y,\vec{t})(Z,\vec{s})$ for all
     $\vec{s},\vec{t} \in \States$ such that 
     $\gamma_i(\vec{s}(i,r),Z) = \vec{t}(i,r)$ for all 
     $1\leq i \leq k$ and $r \in Q$.
  \end{itemize}
  So, the $\Delta$-automata $\A_1,\cdots,\A_k$ are simulated ``on-the-fly''
  by storing the vector of current states directly in the stack. Hence,
  the information whether a given $\A_i$ accepts the current
  configuration is available in the topmost stack symbol.
  For every $1 \leq i \leq k$, the underlying set $G_i$ of $\conf_i'$
  (see Definition~\ref{def-simple}) is defined by 
  \[
    G_i = \{p(X,\vec{s}) \mid \gamma_i(\vec{s}(i,p),X) \in \Acc_i\} 
          \cup \{p\varepsilon \mid p\varepsilon \in \conf_i\}
  \]
  The function $\G$ is defined by $\G(p\varepsilon) = p\varepsilon$, and
  $\G(p X_1 \cdots X_k) = p (X_1,\vec{s}_1) \cdots (X_k,\vec{s}_k)$,
  where $\vec{s}_k(i,q) = q$, and 
  $\vec{s}_j(i,q) =  \gamma_i(\vec{s}_{j+1}(i,q),X_{j+1})$ for all
  $1 \leq j <k$. It follows immediately from the definition of $\delta'$
  and $\Prob'$ that the parts of $\T_{\Delta}$ and $\T_{\Delta'}$ which
  are reachable from $p\alpha$ and $\G(p\alpha)$ are isomorphic (for
  every $p\alpha \in \conf(\Delta)$).

  Let $\conf \subseteq \conf(\Delta')$ be a regular set of 
  configurations.
  Since some configurations of $\conf$ can be ``inconsistent'' 
  in the sense that the vectors of states that are stored together with 
  the original stack symbols do not correspond to a valid computation of 
  the $\A_i$ automata, the set $\G^{-1}(\conf)$ is not a simple projection
  of $\conf$ ``forgetting'' the vectors of states from the stack 
  symbols. Fortunately, $\G(\conf(\Delta))$ is (obviously) a regular
  set, so we can construct a $\Delta'$-automaton recognizing
  the set $\conf \cap \G(\conf(\Delta))$ and apply the mentioned
  projection. 
\end{proof}

\section{Random Walks on pPDA Graphs}
\label{sec-random}

In this section we address the following problem. Let
$\Delta$ be a pPDA, let $p_1\alpha_1$ be an initial configuration,
let $\conf_1, \conf_2$ be two simple sets of configurations, and let 
$\rho$ be a threshold probability. Is the probability of
executing a run $p_1\alpha_1;p_2\alpha_2;p_3\alpha_3 \cdots$ that satisfies
$\conf_1 \U \conf_2$, denoted by $\calP(p_1\alpha_1,\conf_1 \U \conf_2)$,
at least $\rho$? We show that the problem is decidable.

The plan of the section is as follows. First, we show 
in Lemma \ref{lem-prob-def} that $\calP(p_1\alpha_1,\conf_1 \U \conf_2)$
is equal to a polynomial expression in the following probabilities:
\begin{itemize}
\item Let $pX$ be an initial configuration (notice that there is only one symbol on the stack), and let $q$ be a control state $q$. The probability of 
reaching $q\varepsilon$ visiting only configurations of $\conf_1 \smallsetminus \conf_2$ along the way is denoted by $\Pro{pXq}$
\item   Let $pX$ be an initial configuration and let $\tau$ be a 
threshold probability.  The probability
of reaching some configuration of
$\conf_2$ with nonempty stack, visiting only configurations of $\conf_1$
along the way, is denoted by $\Pro{pX\bullet}$.
\end{itemize}
\noindent Second, in Theorem \ref{thm-least-fix}, we show that
the probabilities $\Pro{pXq}$ and $\Pro{pX\bullet}$ are the least solution
of a system of quadratic equations. So our original problem reduces
to determining whether a polynomial expression on this least solution 
has at least the value $\rho$. Finally, we observe in
Theorem \ref{thm-arithmetic} that this question 
can be reduced to deciding the truth of a formula in the first-order
arithmetic of the reals (i.e., in the theory $(\mathbb{R},+,*,\leq)$).
Since this theory is known to be decidable \cite{Tarski:reals-arithmetic},
our original question is decidable.

For the rest of this section, let us fix a pPDA 
$\Delta = (Q,\Gamma,\delta,\Prob)$ and two simple sets
$\conf_1,\conf_2 \subseteq \conf(\Delta)$. Let 
$G_1,G_2 \subseteq Q \times (\Gamma \cup \{\varepsilon\})$ be the 
sets associated to $\conf_1,\conf_2$ in the sense of 
Definition~\ref{def-simple}.

\begin{defi}
\label{def-notation}
To simplify our notation, we adopt the following conventions:
\begin{itemize}
\item For each $\conf \subseteq \conf(\Delta)$, let 
   $\conf^\bullet = \conf \smallsetminus (Q {\times} \{\varepsilon\})$. 
   Observe that if $\conf$ is simple, then so is $\conf^\bullet$.
\item For every $\conf \subseteq \conf(\Delta)$ and every 
   $\beta \in \Gamma^*$, the symbol $\conf\beta$ denotes the set
   $\{p\alpha\beta \mid p\alpha \in \conf\}$.
\item For all $p,q \in Q$ and $X \in \Gamma$, we use 
   $\Pro{pXq}$ to abbreviate 
   $\calP(pX,\conf_1 {\smallsetminus} \conf_2 \U \{q \varepsilon\})$, 
   and  $\Pro{pX\bullet}$ 
   to abbreviate 
   $\calP(pX, \conf_1 \U \conf^\bullet_2)$.
\item Let $A$ be a set of finite paths which end in the same state $t$, and
   let $B$ a set of finite or infinite paths that start in $t$. Then
   the symbol $A \odot B$ denotes the set of paths 
   $\{v;w \mid v \in A, t;w \in B\}$.
\end{itemize}
\end{defi}

The proof of Lemma~\ref{lem-prob-def}. our first milestone,
requires the following two auxiliary results:

\begin{lem}
\label{lem-concat-path} 
Let $\T = (S,\tran{},\Prob)$ be a probabilistic transition system.
Let $s,t \in S$ and $\conf_1,\conf_2 \subseteq S$. Further, let 
$A = \fpath(s,(\conf_1 {\smallsetminus} \conf_2) \U \{t\})$ and
$B = \fpath(t, \conf_1 \U \conf_2)$. Then 
\[
  \sum_{w \in A \odot B} \calP(\run(w)) = 
  \sum_{w \in A} \calP(\run(w)) \cdot \sum_{w \in B} \calP(\run(w)).
\]
\end{lem}
\begin{proof} Immediate.
\end{proof}

\begin{lem}
\label{lem-tail}
For all $p\alpha \in \conf(\Delta)$ and $\beta \in \Gamma^*$ we have that
$\calP(p\alpha,\conf_1 \U \conf_2)$ is equal to 
$\calP(p\alpha\beta,\conf_1^\bullet\beta \U \conf_2\beta)$.
\end{lem}
\begin{proof}
  For every finite path $w = p_1\alpha_1; \cdots; p_n\alpha_n$ of 
  $\fpath(p\alpha)$, let $w^{+\beta}$ denote the finite path
  $p_1\alpha_1\beta; \cdots; p_n\alpha_n\beta$
  of $\fpath(p\alpha\beta)$. Realize that 
  $\calP(\run(w)) = \calP(\run(w^{+\beta}))$, because $w$ and $w^{+\beta}$
  execute the same transitions. One can easily verify that 
  $w \in \fpath(p\alpha,\conf_1 \U \conf_2)$ if{}f 
  $w^{+\beta} \in \fpath(p\alpha\beta,\conf_1^\bullet\beta \U \conf_2\beta)$.
  From this we get
  \begin{eqnarray*}
   \calP(p\alpha,\conf_1 \U \conf_2) & = & 
     \sum_{w \in \fpath(p\alpha,\conf_1 \U \conf_2)} \calP(\run(w))\\
   & = &
     \sum_{w \in \fpath(p\alpha\beta,\conf_1^\bullet\beta \U \conf_2\beta)}
     \calP(\run(w))\\
   & = & \calP(p\alpha\beta,\conf_1^\bullet\beta \U \conf_2\beta)
  \end{eqnarray*}
\end{proof}

Now we show how to compute $\calP(p X_1 \cdots X_n, \conf_1 \U \conf_2)$
from the finite family of all $\Pro{pXq}$, $\Pro{pX\bullet}$ probabilities.
First, realize that
\begin{eqnarray*}
  \calP(p X_1 \cdots X_n, \conf_1 \U \conf_2) & = & 
  \Pro{p X_1 \bullet} + \sum_{q \in Q} \Pro{pX_1q} \cdot 
  \calP(q X_2 \cdots X_n, \conf_1 \U \conf_2)
\end{eqnarray*}
The meaning of this equation is intuitively clear. If we repeatedly 
expand the probabilities of the form 
$\calP(q X_j \cdots X_n, \conf_1 \U \conf_2)$ in the above equation
(until $j$ becomes $n$), we obtain the equation presented
in the following lemma: 

\begin{lem}
\label{lem-prob-def}
  For each $p X_1 \cdots X_n \in \conf(\Delta)$ where $n \geq 0$
  we have that  $\calP(p X_1 \cdots X_n, \conf_1 \U \conf_2)$ is equal to
  \[
    \sum_{i=1}^{n} \sum_{\substack{(q_1,\cdots,q_i) \in Q^i \\ \text{where }
        p=q_1}}
     \Pro{q_i X_i \bullet} \cdot
     \prod_{j=1}^{i-1} 
     \Pro{q_j X_j q_{j+1}}
     \ +\  \sum_{\substack{(q_1,\cdots,q_{n+1}) \in Q^{n+1} \\ 
                          \text{where } p=q_1 \text{ and } 
                          q_{n{+}1}\varepsilon \in \conf_2}}
          \prod_{j=1}^{n} 
     \Pro{q_j X_j q_{j+1}}    
  \]
  with the convention that empty sum is equal to $0$ and empty product
  is equal to $1$.  
\end{lem}
\begin{proof}
  By induction on $n$. For $n=0$ we have that 
  $\calP(p \varepsilon, \conf_1 \U \conf_2)$ is equal either to $1$ or $0$,
  depending on whether $p \varepsilon$ belongs to $\conf_2$ or not, resp.
  Now let $n \geq 1$, and let $\beta$ denote the sequence $X_2 \cdots X_n$. 
  The set $\run(p X_1 \beta, \conf_1 \U \conf_2)$ is equal to
  \[
    \biguplus_{w \in \fpath(pX_1\beta, \conf_1 \U \conf_2)} \run(w)
  \]
  Let $\conf' = \{q\alpha\beta \mid q \in Q, 
  \alpha \in \Gamma^+ \}$. We have that
  \begin{tabbing}
  \hspace*{0em} \= \hspace*{1em} \= \kill
    $\fpath(pX_1\beta, \conf_1 \U \conf_2)\ =
        \fpath(pX_1\beta,
          \conf_1 {\cap} \conf' \U \conf_2 {\cap} \conf') \ \ \uplus$\\[1ex]
    \>  \>
       $\displaystyle\biguplus_{q \in Q} \fpath(pX_1\beta,
       (\conf_1{\smallsetminus} \conf_2) {\cap} \conf' 
           \U\{q\beta\})\odot
       \fpath(q\beta, \conf_1 \U \conf_2)$
  \end{tabbing}
  Now observe that for every \emph{simple} set 
  $\conf \subseteq \conf(\Delta)$ we have that
  $\conf \cap \conf' = \conf^\bullet \beta$. Hence, the above equation can
  be rewritten as follows:
  \begin{tabbing}
  \hspace*{0em} \= \hspace*{1em} \= \kill
    $\fpath(pX_1\beta, \conf_1 \U \conf_2)\ =
        \fpath(pX_1\beta,
          \conf_1^\bullet \beta \U \conf_2^\bullet \beta) \ \ \uplus$\\[1ex]
    \>  \>
       $\displaystyle\biguplus_{q \in Q} \fpath(pX_1\beta,
       (\conf_1{\smallsetminus} \conf_2)^\bullet \beta 
           \U\{q\beta\})\odot
       \fpath(q\beta, \conf_1 \U \conf_2)$
  \end{tabbing}
  Using Lemma~\ref{lem-tail} and Lemma~\ref{lem-concat-path}, we obtain that 
  \begin{tabbing}
  \hspace*{0em} \= \hspace*{1em} \= \kill
    $\calP(pX_1\beta, \conf_1 \U \conf_2)\ =
        \calP(pX_1, \conf_1 \U \conf_2^\bullet) \ \ +$\\[1ex]
    \>  \>
       $\sum_{q \in Q} \calP(pX_1\beta,
       (\conf_1{\smallsetminus} \conf_2) \U\{q\beta\}) \cdot
       \calP(q\beta, \conf_1 \U \conf_2)$
  \end{tabbing}
  This can also be written as \[\calP(pX_1\beta, \conf_1 \U \conf_2)\ =
  \Pro{p X_1 \bullet} + \sum_{q \in Q} \Pro{p X_1 q} \cdot 
      \calP(q\beta,\conf_1 \U \conf_2)\]
  Now it suffices to apply induction hypothesis to
  $\calP(q\beta,\conf_1 \U \conf_2)$ and restructure the
  resulting expression.
\end{proof}

Now we show that the probabilities $\Pro{pXq}$, $\Pro{pX\bullet}$ form
the least solution of an effectively constructible system of 
quadratic equations. This can be seen as a generalization of a similar 
result for finite-state systems 
\cite{HJ:logic-time-probability-FAC,CY:probab-verification-JACM}. 
In the finite-state case, the
equations are linear and can be further modified so that they have
a \emph{unique} solution (which is then computable, e.g., 
by Gauss elimination). In the case of pPDA, the equations are not 
linear and 
cannot be generally solved by analytical methods. The question whether
the equations can be further modified so that they have a unique
solution is left open; we just note that the method used for
finite-state systems is insufficient (this is demonstrated by
Example~\ref{exa-ber}).

Let $\V = \{\Var{pXq},\Var{pX\bullet} \mid p,q \in Q, X \in \Gamma\}$
be a set of ``variables''.
Let us consider the system of recursive equations constructed as follows:
\begin{itemize}
\item if $pX \not\in G_1 {\smallsetminus} G_2$, then $\Var{pXq} = 0$ for 
   each $q \in Q$; otherwise, we put
   \[
     \Var{pXq} =  \sum_{pX \tran{x} rYZ} 
                  x \cdot \sum_{t \in Q} \Var{rYt} \cdot \Var{tZq}
                 ~~~ + ~~~ \sum_{pX \tran{x} rY}
                  x \cdot \Var{rYq}
                 ~~~ + ~~~  \sum_{pX \tran{x} q \varepsilon} x   
   \]
\item if $pX \in G_2$, then $\Var{pX\bullet} = 1$; if 
    $pX \not\in G_1 \cup G_2$, then $\Var{pX\bullet} = 0$; otherwise
    we put
    \[
      \Var{pX\bullet}  =  \sum_{pX \tran{x} rYZ} 
                  x \cdot (\Var{rY\bullet} ~~~+~~~ 
                           \sum_{t \in Q} \Var{rYt} \cdot \Var{tZ\bullet})
             ~~~+~~~  \sum_{pX \tran{x} rY}
                  x \cdot \Var{rY\bullet}
    \]
\end{itemize}
The intuition behind these equations is easy to understand. For the sake
of simplicity, assume $G_1 = Q \times \Gamma$ and $G_2 = \emptyset$ 
(this corresponds to $\conf_1 = \conf(\Delta)$ and $\conf_2 = \emptyset$). In this case, 
we only have the two ``long'' equations. Consider the
first one, the intuition for the second one being similar. In order to reach $
q\varepsilon$ from $pX$, the pPDA must make at least one move. Since we assume than 
the transitions $pX \tran{x} q\alpha$
of a pPDA satisfy $|\alpha|\leq 2$, here are three possible kinds of moves: 
moves that increase the stack length by one, moves that do not change the 
stack length, and moves that decrease
the stack length. The three summands in the equations correspond to these 
three kinds of moves. Since no transition can be executed when the stack is empty, 
the only way to reach $q\varepsilon$ by means of a length-decreasing
move is to apply a transition $pX \tran{x} q\varepsilon$, if it exists (third summand). 
If the first transition is length-keeping, i.e., of the form
$pX \tran{x} rY$, then, after the transition, we must reach $q\varepsilon$ 
from $rY$ (second summand).
Finally, if the first transition is of the form $pX \rightarrow rYZ$,
then the pPDA  must first go from $rYZ$ to some configuration $tZ$ along a path of 
configurations having with $Z$ as bottom stack symbol, and then 
from $tZ$ to $q\varepsilon$. Intuitively (see the next theorem for the formal proof), 
the probability of reaching $tZ$ 
from $rYZ$ along such a path is equal to the probability of reaching 
$t\varepsilon$ from $rY$, and so we get the first summand.

For given $t \in [0,1]^{|\V|}$, $p,q \in Q$, and $X \in \Gamma$
we use $\Var{pXq}_t$ and $\Var{pX\bullet}_t$
to denote the component of $t$ which corresponds to the variable
$\Var{pXq}$ and $\Var{pX\bullet}$, respectively.
The above defined system of equations determines a unique operator 
$\F : [0,1]^{|\V|} \rightarrow [0,1]^{|\V|}$ where $\F(t)$ is the tuple
of values obtained by evaluating the right-hand sides of the equations
where all $\Var{pXq}$ and $\Var{pX\bullet}$ are substituted with
$\Var{pXq}_t$ and $\Var{pX\bullet}_t$, respectively.
 
\begin{thm}
\label{thm-least-fix}
  The operator $\F$ has the least fixed-point~$\mu$. Moreover,
  for all $p,q \in Q$ and $X \in \Gamma$ we have that
  $\Var{pXq}_\mu = \Pro{pXq}$ and $\Var{pX\bullet}_\mu = \Pro{pX\bullet}$. 
\end{thm}
\begin{proof} Since $\F$ is monotonic and continuous, it has
the least fixed point $\mu = \bigvee_{k=0}^\infty \F^k(\vec{0})$,
where $\vec{0}$ is the tuple of zeros.
One can readily check that the tuple $\pi$ of all $\Pro{pXq}$ and 
$\Pro{pX\bullet}$ probabilities forms a solution of the above system; 
this is done just by partitioning the associated sets of runs into 
appropriate disjoint subsets similarly as in the proof of
Lemma~\ref{lem-prob-def}. Hence, $\mu \leq \pi$. To prove that
also $\pi \leq \mu$, we approximate the $\Pro{pXq}$ and 
$\Pro{pX\bullet}$ probabilities in the following way: For each 
$k \in \Nset_0$ we define
\begin{itemize}
  \item $\displaystyle\Pro{pXq}^k = \sum_{w \in 
     \fpath^k(pX, \conf_1 {\smallsetminus} \conf_2 \U \{q\varepsilon\})} 
     \calP(\run(w))$ \smallskip

  \item $\displaystyle\Pro{pX\bullet}^k = \sum_{w \in 
     \fpath^k(pX, \conf_1 \U \conf_2^\bullet)} \calP(\run(w))$ 
\end{itemize}
Let $\pi^k$ be the tuple of all $\Pro{pXq}^k$ and $\Pro{pX\bullet}^k$
probabilities. Clearly $\pi = \lim_{k \rightarrow \infty} \pi^k$.
By induction on $k$ we prove that $\pi^k \leq \mu$ for each $k \in \Nset_0$, 
hence also $\pi \leq \mu$ as needed.

The base case ($k=0$) follows immediately. We show that if
 $\Pro{pXq}^k \leq \Var{pXq}_\mu$ and $\Pro{pX\bullet}^k \leq \Var{pXq}_\mu$,
then also $\Pro{pXq}^{k+1} \leq \Var{pXq}_\mu$ and 
$\Pro{pX\bullet}^{k+1} \leq \Var{pXq}_\mu$. If 
$pX \not\in G_1 {\smallsetminus} G_2$, then $\Pro{pXq}^{k+1} = 
\Var{pXq}_\mu = 0$. Otherwise, by applying the definitions we obtain
\begin{eqnarray*}
   \Pro{pXq}^{k+1} & = & 
   \sum_{pX \tran{x} rYZ} 
      x \cdot \sum_{w \in \fpath^k(rYZ,\conf_1 {\smallsetminus} \conf_2 \U
              \{q\varepsilon\})} \calP(\run(w))\\
   & + & \sum_{pX \tran{x} rY}
         x \cdot \sum_{w \in \fpath^k(rY,\conf_1 {\smallsetminus} 
         \conf_2 \U \{q\varepsilon\})} \calP(\run(w))\\
   & + & \sum_{pX \tran{x} q \varepsilon} x
\end{eqnarray*}
and
\[ 
  \Var{pXq}_\mu \ = \ 
    \sum_{pX \tran{x} rYZ} 
           x \cdot \sum_{t \in Q} \Var{rYt}_\mu \cdot \Var{tZq}_\mu
   ~~~+~~~ \sum_{pX \tran{x} rY} x \cdot \Var{rYq}_\mu
   ~~~+~~~ \sum_{pX \tran{x} q \varepsilon} x
\]
Since 
\[ \sum_{w \in \fpath^k(rY,\conf_1 {\smallsetminus} 
   \conf_2 \U \{q\varepsilon\})} \calP(\run(w)) \quad = \quad \Pro{rYq}^k,\] 
we have
\[ \sum_{w \in \fpath^k(rY,\conf_1 {\smallsetminus} 
  \conf_2 \U \{q\varepsilon\})} \calP(\run(w)) \quad \leq \quad \Var{rYq}_\mu\]
by induction hypothesis. Further, \[\sum_{pX \tran{x} rYZ} 
      x \cdot \sum_{w \in \fpath^k(rYZ,\conf_1 {\smallsetminus} \conf_2 \U
              \{q\varepsilon\})} \calP(\run(w))\] is surely
bounded by \[\sum_{pX \tran{x} rYZ} 
           x \cdot \sum_{t \in Q} \Pro{rYt}^k \cdot \Pro{tZq}^k,\] which
is bounded by \[\sum_{pX \tran{x} rYZ} 
           x \cdot \sum_{t \in Q} \Var{rYt}_\mu \cdot \Var{tZq}_\mu\] by
induction hypothesis. To sum up, we have that 
$\Pro{pXq}^{k+1} \leq \Var{pXq}_\mu$. The inequality 
$\Pro{pX\bullet}^{k+1} \leq \Var{pX\bullet}_\mu$ is proved similarly.
\end{proof}

\begin{exa}
\label{exa-ber}
Let us consider the pBPA system $\Delta$ of Fig.~\ref{fig-walk}, and let
$\conf_1 = \Gamma^*$, $\conf_2 = \{Z\}$. Then we
obtain the following system of equations (since $\Delta$ has only
one control state $p$, we write $\Var{X,\bullet}$ and $\Var{X,\varepsilon}$
instead of $\Var{pX\bullet}$ and $\Var{pXp}$, resp.):
\begin{eqnarray*}
  \Var{Z,\bullet} & = & 1\\
  \Var{Z,\varepsilon} & = &
     x \Var{I,\varepsilon} \Var{Z,\varepsilon}\ +\ 
     (1{-}x) \Var{D,\varepsilon} \Var{Z,\varepsilon}\\ 
  \Var{I,\bullet} & = & x(\Var{I,\bullet}\ +\  
     \Var{I,\varepsilon} \Var{I,\bullet})\\
  \Var{I,\varepsilon} & = & x \Var{I,\varepsilon}\Var{I,\varepsilon}\ +\ 
     1{-}x\\
  \Var{D,\bullet} & = & (1{-}x)(\Var{D,\bullet}\ +\  
     \Var{D,\varepsilon} \Var{D,\bullet})\\
  \Var{D,\varepsilon} & = & (1{-}x) \Var{D,\varepsilon}\Var{D,\varepsilon}\ 
     +\ x
\end{eqnarray*}
As the least solution we obtain the probabilities 
$\Pro{Z,\bullet} = 1$, $\Pro{Z,\varepsilon} = 0$, 
$\Pro{I,\bullet} = 0$, $\Pro{I,\varepsilon} = \min\{1,(1{-}x)/x\}$,
$\Pro{D,\bullet} = 0$, $\Pro{D,\varepsilon} = \min\{1,x/(1{-}x)\}$.
By applying Lemma~\ref{lem-prob-def} we further obtain that, e.g.,
$\calP(IIZ, \conf_1 \U \conf_2) = 
\Pro{I,\bullet} + \Pro{I,\varepsilon} \cdot (\Pro{I,\bullet} + 
\Pro{I,\varepsilon} \cdot \Pro{Z,\bullet}) = 
\min \{1, (1{-}x)^2/x^2\}$. \hfill$\Box$
\end{exa}

In Example~\ref{exa-ber} it is possible to compute a closed
form for the least solution of the system
of equations, but in general this is not true. However,
many important properties of the least solution are decidable, because the
decision problem can be reduced to the problem of deciding the truth of a formula
in the first-order theory of the reals. 
For our purposes, it suffices to consider the class of properties 
defined in the next theorem.

\begin{thm}
\label{thm-arithmetic}
  Let $\mathit{Const} = \mathbb{Q} \cup 
  \{\Pro{pXq},\Pro{pX\bullet} \mid p,q \in Q \text{ and } X \in \Gamma\}$,
  where $\mathbb{Q}$ is the set of all rational constants. 
  Let $E_1,E_2$ be expressions built over $\mathit{Const}$ using
  `$\cdot$' and `$+$', and let ${\sim} \in \{<,=\}$. It is decidable whether
  $E_1 \sim E_2$.
\end{thm}
\begin{proof}
  We show that, due to Theorem~\ref{thm-least-fix}, $E_1 \sim E_2$ is
  effectively expressible as a closed formula of $(\mathbb{R},+,*,\leq)$.
  Hence, the theorem follows from the decidability of first-order 
  arithmetic of reals \cite{Tarski:reals-arithmetic}.
  
  For all $p,q \in Q$ and $X \in \Gamma$, let
  $x(pXq)$, $x(pX\bullet)$, $y(pXq)$, and $y(pX\bullet)$ be first order
  variables, and let $\vec{X}$ and $\vec{Y}$ be the vectors of all
  $x(pXq)$, $x(pX\bullet)$, and $y(pXq)$, $y(pX\bullet)$ variables,
  respectively. Let us consider the formula $\Phi$ constructed as follows:
  \begin{tabbing}
  \hspace*{1em} \= \hspace*{1em} \= \kill
   $\exists \vec{X}~:~ \vec{0} \leq \vec{X} \leq \vec{1} \quad \wedge\quad 
      \vec{X} = \F(\vec{X})$\\[1ex]
   \> $\wedge$ \>
   $(\forall \vec{Y} ~:~ (\vec{0} \leq \vec{Y} \leq \vec{1} \ \wedge\ 
     \vec{Y} = \F(\vec{Y})) \ \Rightarrow\  
     \vec{X} \leq \vec{Y}))$\\[1ex]
   \> $\wedge$ \> $E_1[\vec{X}/\pi] \sim E_2[\vec{X}/\pi]$
   \end{tabbing}
   Observe that the conditions $\vec{X} = \F(\vec{X})$ and 
   $\vec{Y} = \F(\vec{Y})$ are expressible only using multiplication,
   summation, and equality. The expressions $E_1[\vec{X}/\pi]$ and  
   $E_2[\vec{X}/\pi]$ are obtained from $E_1$ and $E_2$ by substituting
   all $\Pro{pXq}$ and $\Pro{pX\bullet}$ with $x(pXq)$ and $x(pX\bullet)$,
   respectively. It follows immediately that $E_1 \sim E_2$ iff
   $\Phi$ holds.
\end{proof}

\begin{figure}[http]
{\small
\begin{tabbing}
\hspace*{1.5em} \=  \hspace*{1em} \= \hspace*{1em} \= \hspace*{1em} \=\kill
\textbf{Input:} $pX \in \conf(\Delta)$, 
   $0 < \lambda < 1$\\
\textbf{Output:} $\calP^\ell$, $\calP^u$\\[1ex]
\texttt{1:} 
\> $\calP^\ell := 0;$ $\calP^u := 1;$\\
\texttt{2:}
\> \kw{for } $i=1$ \kw{ to } $\lceil -\log_2 \lambda  \rceil$\\
\texttt{3:}
\>\> \kw{if } $\Pro{pX\bullet} + 
  \sum_{q\varepsilon \in \conf_2} \Pro{pXq} \geq (\calP^u - \calP^\ell)/2$\\
\texttt{4:}
\>\>\> \kw{then } $\calP^\ell := (\calP^u - \calP^\ell)/2$\\
\texttt{5:}
\>\>\> \kw{else } $\calP^u := (\calP^u - \calP^\ell)/2$\\
\texttt{6:}
\>\> \kw{fi}
\end{tabbing}}
\caption{Computing $\calP^\ell,\calP^u$}
\label{fig-alg-random}
\end{figure}

An immediate consequence of Theorem~\ref{thm-arithmetic} is the following:

\begin{thm}
  Let $p\alpha \in \conf(\Delta)$, $\varrho \in \mathbb{Q} \cap [0,1]$, 
  ${\sim} \in \{{\leq},{<},{\geq},{>}\}$ and $0 < \lambda < 1$. 
  It is decidable whether $\calP(p\alpha, \conf_1 \U \conf_2) \sim \varrho$.
  Moreover, there effectively exist rational numbers $\calP^\ell,\calP^u$
  such that $\calP^\ell \leq \calP(p\alpha, \conf_1 \U \conf_2) \leq \calP^u$
  and $\calP^u - \calP^\ell \leq \lambda$.
\end{thm}
\begin{proof}
  We can assume w.l.o.g.\ that $\alpha = X$ for some $X \in \Gamma$.
  Note that
  $\calP(pX, \conf_1 \U \conf_2) \sim \varrho$ iff
  $\Pro{pX\bullet} + 
  \sum_{q\varepsilon \in \conf_2} \Pro{pXq} \sim \varrho$ by
  Lemma~\ref{lem-prob-def}. Hence, we can apply
  Theorem~\ref{thm-arithmetic}.
  The numbers $\calP^\ell,\calP^u$ are computable, e.g., by the
  algorithm of Fig.~\ref{fig-alg-random}. 
\end{proof}

\section{Model Checking PCTL for pPDAs}
\label{sec-PCTL}

In this section we study the model-checking problem for PCTL
formulas with regular valuations and pPDA.

\subsection{Qualitative Fragment of PCTL}
\label{sec-PCTL-quality}

We give a model checking algorithm for the qualitative
fragment of PCTL, i.e., for the fragment in which only 
$0$ and $1$ are allowed as probability thresholds.

Recall that in order to check if a CTL formula $\varphi$ holds of a finite
state system  we first recursively compute the sets of states that satisfy 
the subformulas of $\varphi$ lying right below $\varphi$ in the syntax tree, and then 
we apply a semantic operator that gets these sets of states as inputs and produces the
set of states satisfying $\varphi$ as output. In the case of a PDA (no probabilities), 
these sets of states (they are now sets of configurations) 
can be infinite. Therefore, in order to apply a similar
algorithm it is necessary to prove that the sets have a finite representation.
This was done in \cite{BEM:PDA-reachability}: It was shown that in the case of 
regular valuations the sets are always regular, and so can be finitely
represented by, say, finite automata. In this section we prove that
the same property also holds for pPDA and for the qualitative fragment of PCTL,
and that the constructions showing the regularity of the sets are effective.

By Lemma~\ref{lem-reg-sim}, we only need to show that if the sets of configurations
satisfying the subformulas of $\varphi$ are simple, then 
the set of configurations satisfying $\varphi$ is regular. We need to consider four
cases, corresponding to formulas of the form $\X^{= 0} \varphi$, 
$\X^{= 1} \varphi$, $\varphi_1 \U^{=0} \varphi_2$, and $\varphi_1 \U^{=1} \varphi_2$.
they are dealt with in Lemma \ref{lem-qualit-X}, Lemma \ref{lem-qualit-U1}, and
Lemma \ref{lem-qualit-U0}.

For the rest of this section we fix a pPDA $\Delta = (Q,\Gamma,\delta,\Prob)$. 

\begin{lem}
\label{lem-qualit-X}
Let $\conf \subseteq \conf(\Delta)$ be a simple set. The sets
$\{p\alpha \in \conf(\Delta) \mid \calP(p\alpha, \X \conf) = 1\}$ and
$\{p\alpha \in \conf(\Delta) \mid \calP(p\alpha, \X \conf) = 0\}$ are
effectively regular.
\end{lem}
\begin{proof}
Follows immediate from the fact that $p\alpha$ has only finitely many successors
in the probabilistic transition system associated to $\Delta$..
\end{proof}

\begin{lem}
\label{lem-qualit-U1}
Let $\conf_1,\conf_2  \subseteq \conf(\Delta)$ be simple sets.
The set $\{p\alpha \in \conf(\Delta) \mid 
\calP(p\alpha, \conf_1 \U \conf_2) = 1\}$ is effectively regular.
\end{lem}
\begin{proof}
  Let $R(pX) = \{q \in Q \mid [pXq] > 0\}$ for all $p \in Q$, $X \in \Gamma$. 
  For each $i \in \Nset_0$ we define the set $S_i \subseteq \conf(\Delta)$  
  inductively as follows:
  \begin{itemize}
  \item $S_0  = \{q \varepsilon \mid q \varepsilon \in \conf_2\} \cup
     \{qX\alpha \mid \Pro{qX\bullet} = 1, \alpha \in \Gamma^*\}$
  \item $S_{i+1} = \{pX\beta \mid [pX\bullet] + \sum_{q \in R(pX)} [pXq] =1
    \text{ and } \forall q \in R(pX) : q\beta \in S_i\}$
  \end{itemize}
  Using Lemma~\ref{lem-prob-def}, we can easily check that  
  $\bigcup_{i=0}^\infty S_i = \{p\alpha \in \conf(\Delta) \mid 
    \calP(p\alpha, \conf_1 \U \conf_2) =1\}$.
  To see that 
  the set $\bigcup_{i=0}^\infty S_i$ is effectively regular, for each
  $p \in Q$ we construct a finite automaton $\M_p$ such that
  $L(\M_p) = \{ \alpha \in \Gamma^* \mid 
  p\alpha \in \bigcup_{i=0}^\infty S_i\}$. A $\Delta$-automaton $\A$
  recognizing the set $\bigcup_{i=0}^\infty S_i$ can then be constructed
  using standard algorithms of automata theory (in particular, note
  that regular languages are effectively closed under reverse). The states
  of $\M_p$ are all subsets of $Q$, $\{p\}$ is the initial state,
  $\Gamma$ is the input alphabet, the final states are those $T \subseteq Q$
  where for every $q \in T$ we have that $q\varepsilon \in \conf_2$
  (in particular, note that $\emptyset$ is a final state), and the transition
  function is given by $T \tran{X} U$ if{}f for every $q \in T$ we have
  that $[qX\bullet] + \sum_{r \in R(qX)} [qXr] =1$ and
  $U = \bigcup_{q \in T} R(qX)$. Note that $\emptyset \tran{X} \emptyset$
  for each $X \in \Gamma$. The definition of $\M_p$ is effective 
  due to Theorem~\ref{thm-arithmetic}. It is straightforward to check that 
  $L(\M_p) = \{ \alpha \in \Gamma^* \mid  
  p\alpha \in \bigcup_{i=0}^\infty S_i\}$.
 \end{proof}

\begin{lem}
\label{lem-qualit-U0}
Let $\conf_1,\conf_2  \subseteq \conf(\Delta)$ be simple sets.
The set $\{p\alpha \in \conf(\Delta) \mid 
\calP(p\alpha, \conf_1 \U \conf_2) = 0\}$ is effectively regular.
\end{lem}
\begin{proof}
  Let $R(pX) = \{q \in Q \mid [pXq] > 0\}$ for all $p \in Q$, $X \in \Gamma$. 
  For each $i \in \Nset_0$ we define the set $S_i \subseteq \conf(\Delta)$  
  inductively as follows:
  \begin{itemize}
  \item $S_0  =  \{q \varepsilon \mid q \varepsilon \not\in \conf_2\}$
  \item $S_{i+1} = \{pX\beta \mid [pX\bullet] = 0
    \text{ and } \forall q \in R(pX) : q\beta \in S_i\}$
  \end{itemize}
  The fact
  $\bigcup_{i=0}^\infty S_i = \{p\alpha \in \conf(\Delta) \mid 
    \calP(p\alpha, \conf_1 \U \conf_2)=0 \}$ follows immediately from
  Lemma~\ref{lem-prob-def}. The set $\bigcup_{i=0}^\infty S_i$ is effectively
  regular, which can be shown by constructing a finite automaton
  $\M_p$ recognizing the set $\{ \alpha \in \Gamma^* \mid  
  p\alpha \in \bigcup_{i=0}^\infty S_i\}$. This construction and the
  rest of the argument are very similar to the ones of the proof
  of Lemma~\ref{lem-qualit-U1}. Therefore, they are not given explicitly.
 \end{proof}

\begin{thm}
\label{thm-PDA-CTL}
Let $\varphi$ be a qualitative PCTL formula and
$\nu$ a regular valuation. The set $\{p\alpha \in \conf(\Delta) \mid
p\alpha \models^\nu \varphi\}$ is effectively regular.
\end{thm}
\begin{proof}
  By induction on the structure of $\varphi$. The cases when 
  $\varphi \equiv \mathtt{tt}$ and $\varphi \equiv a$ follow immediately.
  For Boolean connectives we use the fact that regular sets are closed
  under complement and intersection. The other cases are covered
  by Lemma~\ref{lem-qualit-X},~\ref{lem-qualit-U1},~and~\ref{lem-qualit-U0}.
  Here we also need Lemma~\ref{lem-reg-sim}, because the regular sets
  of configurations must effectively be replaced with simple ones before
  applying Lemma~\ref{lem-qualit-X},~\ref{lem-qualit-U1}, 
  and~\ref{lem-qualit-U0}.
\end{proof}

\subsection{Model Checking PCTL for pBPA Processes}
\label{sec-PCTL-nBPA}

In this section we consider arbitrary PCTL properties with regular valuations, but restrict 
ourselves to pBPA processes. We provide an error-tolerant model-checking algorithm.
Since it is not so 
obvious what is meant by error tolerance in the context of PCTL model checking,
this notion is defined formally. More precisely, we first show that for every
formula there is an equivalent negation-free formula, and then we provide
a definition for negation-free formulas. 

Let $\T = (S, \tran{},\Prob)$ be a probabilistic transition system and
$0 < \lambda < 1$,  let $\varphi$ be a PCTL formula, and 
let $\nu$ be a regular valuation (i.e., for every atomic proposition $a$ 
the set $\nu(a)$ of configurations is regular). We observe that there is a 
negation-free formula
$\varphi'$ and a regular valuation $\nu'$ such that $\cosem \varphi \cfsem^\nu = 
\cosem \varphi' \cfsem^{\nu'}$. 
First, negations can be ``pushed inside'' to atomic propositions
using dual connectives (note that, e.g., 
$\neg (\varphi \U^{\geq \varrho} \psi)$ is
equivalent to $\varphi \U^{< \varrho} \psi$). Moreover, since
regular sets are closed under complement, $\cosem \neg a \cfsem^\nu$ is also 
regular for every 
$a$. We construct $\varphi'$ by replacing each negation
$\neg a$ by a fresh atomic proposition $b$, and we extend $\nu$ to 
$\nu'$ by defining $\nu(b) = \cosem \neg a \cfsem^\nu$.

For every negation-free PCTL formula $\varphi$ and 
valuation $\nu$ 
we define the denotation of $\varphi$ over $\T$ w.r.t.\ $\nu$ with 
\emph{error tolerance $\lambda$}, denoted $\cosem \varphi \cfsem_\lambda^\nu$,
in the same way as $\cosem \varphi \cfsem^\nu$. The only exception
is $\varphi_1 \U^{\sim \varrho} \varphi_2$ where
\begin{itemize}
 \item if ${\sim} \in \{{<}, {\leq}\}$, then
   $\cosem \varphi_1 \U^{\sim \varrho} \varphi_2 \cfsem^\nu_\lambda  = 
     \{ s \in S \mid \calP(s,\cosem \varphi_1\cfsem^\nu_\lambda \U
          \cosem \varphi_2\cfsem^\nu_\lambda) \sim \varrho + \lambda\}$
 \item if ${\sim} \in \{{>}, {\geq}\}$, then
   $\cosem \varphi_1 \U^{\sim \varrho} \varphi_2 \cfsem^\nu_\lambda  = 
     \{ s \in S \mid \calP(s,\cosem \varphi_1\cfsem^\nu_\lambda \U
          \cosem \varphi_2\cfsem^\nu_\lambda) \sim \varrho - \lambda\}$
\end{itemize}
\noindent Notice that every negation-free formula $\varphi$ satisfies
$\cosem \varphi \cfsem^\nu \subseteq \cosem \varphi \cfsem^\nu_\lambda$.

An \emph{error tolerant PCTL model checking algorithm}
is an algorithm which, for each PCTL formula $\varphi$, valuation $\nu$, 
$s \in S$, and $0 < \lambda < 1$, outputs YES/NO so that
\begin{itemize}
\item if $s \in \cosem \varphi \cfsem^\nu$, then the answer is YES;
\item if the answer is YES, then $s \in \cosem \varphi \cfsem^\nu_\lambda$.
\end{itemize}

For the rest of this section, let us fix a pBPA  
$\Delta = (\Gamma,\delta,\Prob)$. Since $\Delta$ has just one (or ``none'')
control state $p$, we write $\Pro{X,\bullet}$ and $\Pro{X,\varepsilon}$
instead of $\Pro{pX\bullet}$ and $\Pro{pXp}$, respectively.

We need the following obvious generalization of Lemma~\ref{lem-qualit-X}
(use the same proof):

\begin{lem}
\label{lem-BPA-X}
Let $\conf \subseteq \conf(\Delta)$ be a simple set, $\varrho \in [0,1]$,
and ${\sim} \in \{\leq,<,\geq,>\}$. The set
$\{\alpha \in \conf(\Delta) \mid \calP(\alpha, \X \conf) \sim \varrho\}$
is effectively regular.
\end{lem}
\begin{proof}
  Immediate.
\end{proof}

\begin{figure}[http]
\begin{tabbing}
\hspace*{1.5em} \=  \hspace*{1em} \= \hspace*{1em} \= \hspace*{1em} \=\kill
\textbf{Input:}  pBPA $\Delta$, $0 < \lambda < 1$\\
\textbf{Output:} $n$, $\kappa$, $\nu$, $[X,\bullet]^\ell$, 
   $[X,\varepsilon]^\ell$, $[X,\bullet]^u$, $[X,\varepsilon]^u$\\[1ex]
\texttt{1:} 
\> $S := \{ X \in \Gamma \mid [X,\varepsilon] \neq 1\};$\\
\texttt{2:}
\> $\nu := 1;$ $n := \infty$;\\
\texttt{3:}
\> \kw{for each } $X \in S$ \kw{ do}\\
\texttt{4:}
\>\> $[X,\varepsilon]^\ell := 0;$ $[X,\bullet]^\ell := 0;$
     $[X,\varepsilon]^u := 1;$ $[X,\bullet]^u := 1;$\\
\texttt{5:}
\> \kw{done}\\
\texttt{6:}
\> \kw{repeat}\\
\texttt{7:}
\>\> \kw{for each } $X \in \Gamma$ \kw{ do}\\
\texttt{8:}
\>\>\> $\text{avg}^\varepsilon := 
   ([X,\varepsilon]^u - [X,\varepsilon]^\ell)/2;$\\
\texttt{9:}
\>\>\> $\text{avg}^\bullet := 
   ([X,\bullet]^u - [X,\bullet]^\ell)/2;$\\
\texttt{10:}
\>\>\> \kw{if } $[X,\varepsilon] \geq \text{avg}^\varepsilon$
   \kw{ then } $[X,\varepsilon]^\ell := \text{avg}^\varepsilon;$\\
\texttt{11:}
\>\>\>\> \kw{else } $[X,\varepsilon]^u := \text{avg}^\varepsilon;$\\
\texttt{12:}
\>\>\> \kw{if } $[X,\bullet] \geq \text{avg}^\bullet$
   \kw{ then } $[X,\bullet]^\ell := \text{avg}^\bullet;$\\
\texttt{13:}
\>\>\>\> \kw{else } $[X,\bullet]^u := \text{avg}^\bullet;$\\
\texttt{14:}
\>\> \kw{done}\\
\texttt{15:}
\>\> $\nu := \nu / 2;$\\
\texttt{16:}
\>\> $\kappa := \max \{[X,\varepsilon]^u \mid X \in S\};$\\
\texttt{17:}
\>\> \kw{if } $\kappa < 1$ \kw{ then } 
  $n := \lceil (\log (\lambda/3) / \log \kappa  \rceil$\\
\texttt{18:}
\> \kw{until } $\kappa < 1$ \kw{ and } 
      $n(\nu + \nu(n+1)(1+\nu)^n) \leq \lambda/3$
\end{tabbing}
\caption{A part of the algorithm for pBPA}
\label{fig-alg-n}
\end{figure}

The following lemma presents the crucial part of the algorithm. This is
the place where we need the assumption that $\Delta$ is a pBPA. 

\begin{lem}
\label{lem-BPA-U}
  Let $\conf_1,\conf_2 \subseteq \conf(\Delta)$ be simple sets. For 
  all $\varrho \in [0,1]$ and $0 < \lambda < 1$ there effectively
  exist $\Delta$-automata $\A^\geq$ and $\A^\leq$ such that for
  all $\alpha \in \conf(\Delta)$ we have that
  \begin{itemize}
  \item if $\calP(\alpha,\conf_1 \U \conf_2) \geq \varrho$ (or
    $\calP(\alpha,\conf_1 \U \conf_2) \leq \varrho$), then 
    $\alpha \in \conf(\A^\geq)$ (or $\alpha \in \conf(\A^\leq)$, 
    respectively.)
  \item if $\alpha \in \conf(\A^\geq)$ (or $\alpha \in \conf(\A^\leq)$),
    then $\calP(\alpha,\conf_1 \U \conf_2) \geq \varrho - \lambda$
    (or $\calP(\alpha,\conf_1 \U \conf_2) \leq \varrho + \lambda$, 
    respectively.) 
  \end{itemize}
\end{lem}
\begin{proof} We describe just the construction of $\A^\geq$ (the 
  $\Delta$-automaton $\A^\leq$ is constructed similarly).
  Let $S = \{X \in \Gamma \mid \Pro{X,\varepsilon} \neq 1\}$. 
For each $\beta \in S^*$ we define the set 
  $\Cl(\beta) = \{\alpha \in \Gamma^* \mid \restrict{\alpha}{S} = \beta\}$,
  where $\restrict{\alpha}{S}$ is the word obtained by deleting in $\alpha$
  all occurrences of symbols in $\Gamma \smallsetminus S$.
  It follows directly from Lemma~\ref{lem-prob-def} that for all 
  $\beta \in S^*$ and $\alpha \in \Cl(\beta)$ we have that
  $\calP(\beta, \conf_1 \U \conf_2) = \calP(\alpha, \conf_1 \U \conf_2)$. 
  Further, for all $n \in \Nseto$ and $\beta \in \bigcup_{i=0}^n S^i$
  we define the set
  \[
     \Gen_n(\beta) = \begin{cases}
                       \Cl(\beta) & \text{if } \alpha \in S^i \wedge i<n\\
                    \{\alpha\alpha' \mid 
                          \alpha \in \Cl(\beta),\alpha' \in \Gamma^* \} &
                       \text{if } \alpha \in S^n
                  \end{cases}
  \]
  We prove that for every $0 < \lambda < 1$ there effectively exist
  $n \in \Nseto$ and $\G \subseteq \bigcup_{i=0}^n S^i$ such that for 
  every $\alpha \in \Gamma^*$ we have that
  \begin{itemize}
  \item if $\calP(\alpha, \conf_1 \U \conf_2) \geq \varrho$, then
     $\alpha \in \bigcup_{\beta \in \G} \Gen_n(\beta)$;
  \item if $\alpha \in \bigcup_{\beta \in \G} \Gen_n(\beta)$, then
     $\calP(\alpha, \conf_1 \U \conf_2) \geq \varrho - \lambda$.
  \end{itemize}
  This suffices for our purposes, because the set 
  $\bigcup_{\beta \in \G} \Gen_n(\beta)$ is clearly recognizable by
  an effectively constructible $\Delta$-automaton $\A^\geq$.

  The crucial part of the algorithm for computing the set $\G$ is
  shown in Fig.~\ref{fig-alg-n}. The algorithm starts by computing the 
  set $S$ (note that $S$ is effectively computable 
  due to Theorem~\ref{thm-arithmetic}). For each $X \in S$, there
  are four rational variables $[X,\varepsilon]^\ell$, 
  $[X,\varepsilon]^u$, $[X,\bullet]^\ell$, and  $[X,\bullet]^u$ whose
  values are lower and upper approximations of the probabilities
  $\Pro{X,\varepsilon}$ and $\Pro{X,\bullet}$, resp. These variables
  are initialized in lines \mbox{3--5} and successively refined in lines
  7--14. Note that the conditions of the \kw{if} statements in lines
  10 and 12 are effective due to Theorem~\ref{thm-arithmetic}.
  The current ``precision'', i.e., the difference between
  the upper and the lower approximation is stored in the rational variable
  $\nu$. The subtle point is the termination condition. First, one
  necessary condition for termination is that 
  $\kappa = \max\{[X,\varepsilon]^u \mid X \in S\}$ becomes less than one.
  This must happen eventually, because $\Pro{X,\varepsilon} < 1$ for 
  every $X \in S$. An important observation is that $\kappa$ can only
  \emph{decrease} by performing the assignment in line~16. This
  means that $n = \lceil \log (\lambda/3) / \log \kappa \rceil$
  also only \emph{decreases} (since both $\lambda$ and $\kappa$
  are less than $1$, we have $\log (\lambda/3) / \log \kappa =
  |\log (\lambda/3)| / |\log \kappa|$; and if $0 < \kappa' < \kappa <1$,
  then $|\log \kappa'| > |\log \kappa|$). Therefore, we eventually 
  find a sufficiently 
  small $\nu$ such that $n(\nu + \nu(n+1)(1+\nu)^n) \leq \lambda/3$.
  
  The output of the algorithm of Fig.~\ref{fig-alg-n} are the (values of the)
  variables $n$, $\nu$, $\kappa$, $[X,\varepsilon]^\ell$, $[X,\varepsilon]^u$, 
  $[X,\bullet]^\ell$, and  $[X,\bullet]^u$ where $X$ ranges over $S$. 
  For each $\beta \in S^*$,
  let $\calP^\ell(\beta, \conf_1 \U \conf_2)$ and
  $\calP^u(\beta, \conf_1 \U \conf_2)$ be the lower and upper approximations
  of $\calP(\beta, \conf_1 \U \conf_2)$ obtained by using the formula
  of Lemma~\ref{lem-prob-def} where  $[X,\varepsilon]^\ell$, 
  $[X,\bullet]^\ell$, and
  $[X,\varepsilon]^u$, $[X,\bullet]^u$ are used instead of 
  $[X,\varepsilon]$, $[X,\bullet]$, respectively. The set $\G$ is 
  constructed as follows:
  \begin{eqnarray*}
    \G & = & \{\beta \in S^i \mid 0 \leq i < n, \calP^u(\beta,\conf_1 \U
               \conf_2) \geq \varrho\}\\
       & \cup &
             \{\beta \in S^n \mid \calP^u(\beta,\conf_1 \U\conf_2)
               \geq \varrho - \lambda/3\}
  \end{eqnarray*}
  To verify that the set $\G$ has the properties mentioned above, we
  need to formulate two auxiliary observations. 
  \begin{itemize}
  \item[(a)] for all $\beta \in S^n$ and $\alpha \in \Gamma^*$ we have that
    \[
      |\calP(\beta, \conf_1 \U \conf_2) - 
      \calP(\beta\alpha,\conf_1 \U \conf_2)| \leq \lambda/3
    \]
    This follows immediately from the following (in)equalities:
    \begin{eqnarray*}
     \calP(\beta\alpha,\conf_1 \U \conf_2) & = &
         \calP(\beta, \conf_1 \U \conf_2^\bullet)
         \ + \  
         \calP(\beta, \conf_1 {\smallsetminus}\conf_2 \U \{\varepsilon\})
         \cdot \calP(\alpha,\conf_1 \U \conf_2)\\[1ex]
     \calP(\beta,\conf_1 \U \conf_2) & \leq &
       \calP(\beta, \conf_1 \U \conf_2^\bullet)\ +\  
       \calP(\beta, \conf_1 {\smallsetminus}\conf_2 \U \{\varepsilon\})\\[1ex]
     \calP(\beta, \conf_1 {\smallsetminus}\conf_2 \U \{\varepsilon\})
      & \leq & \lambda/3
    \end{eqnarray*}
    The first two (in)equalities are obtained just by applying 
    Lemma~\ref{lem-prob-def}. The last one is derived as follows:
    $\calP(\beta, \conf_1 {\smallsetminus}\conf_2 \U \{\varepsilon\})$
    is surely bounded by $\kappa^n$ (by Lemma~\ref{lem-prob-def} and
    the definition of $\kappa$).
    Since $n = \lceil \log (\lambda/3) / \log \kappa  \rceil$, we
    have $n \cdot \log \kappa \leq \log (\lambda/3)$. Hence,
    $\log \kappa^n \leq \log(\lambda/3)$, thus $\kappa^n \leq \lambda/3$.
  \item[(b)] for each $\beta \in \bigcup_{i=0}^n S^i$ we have that
   \[
     \calP^u(\beta,\conf_1 \U \conf_2) - \calP(\beta,\conf_1 \U \conf_2) 
     \leq \lambda/3
   \]
   Let $k = \mathit{length}(\beta)$. A straightforward induction on
   $k$ reveals that 
   $\calP^u(\beta,\conf_1 \U \conf_2) \leq (k+1) \cdot (1+\nu)^k$.
   Now we prove (again by induction on $k$) that
   \[ 
     \calP^u(\beta,\conf_1 \U \conf_2) - \calP(\beta,\conf_1 \U \conf_2)
     \leq k (\nu + \nu(k+1)(1+\nu)^k)
   \]
   The base case (when $k=0)$ is immediate, because 
   $\calP^u(\varepsilon,\conf_1 \U \conf_2) = 
   \calP(\varepsilon,\conf_1 \U \conf_2)$. Now let $\beta = X\beta'$.
   By definition, $\calP^u(X\beta',\conf_1 \U \conf_2) - 
   \calP(X\beta',\conf_1 \U \conf_2)$ is equal to 
   \begin{eqnarray}
     [X,\bullet]^u + [X,\varepsilon]^u \cdot 
     \calP^u(\beta',\conf_1 \U \conf_2)
     \ -\   (\Pro{X,\bullet} + 
     \Pro{X,\varepsilon} \cdot  \calP(\beta',\conf_1 \U \conf_2))
   \label{eqn1}
   \end{eqnarray}
   Since $[X,\bullet]^u \leq \Pro{X,\bullet} + \nu$ and
    $[X,\varepsilon]^u \leq \Pro{X,\varepsilon} + \nu$, the
   expression~(\ref{eqn1}) is bounded by
   \begin{eqnarray}
     \nu \ + \ 
        \Pro{X,\varepsilon}  \cdot (\calP^u(\beta',\conf_1 \U \conf_2)
     - \calP(\beta',\conf_1 \U \conf_2))
     \ +\ \nu \cdot \calP^u(\beta',\conf_1 \U \conf_2)
   \label{eqn3}
   \end{eqnarray}
   By applying induction hypothesis and the facts that 
   $\Pro{X,\varepsilon} \leq 1$ and  
   $\calP^u(\beta,\conf_1 \U \conf_2) \leq (k+1) \cdot (1+\nu)^k$ (see above),
   we obtain that the expression~(\ref{eqn3}) is bounded by
   \[   
      \nu + k (\nu + \nu(k+1)(1+\nu)^k) + \nu(k+1)(1+\nu)^k
   \]
   which is bounded by $(k+1) (\nu + \nu(k+2)(1+\nu)^{k+1})$ as required.
   This finishes the inductive step.

   Since $n(\nu + \nu(n+1)(1+\nu)^n) \leq \lambda/3$ and $k\leq n$,  we have
   $\calP^u(\beta,\conf_1 \U \conf_2) - \calP(\beta,\conf_1 \U \conf_2) 
   \leq k (\nu + \nu(k+1)(1+\nu)^k) \leq \lambda/3$. 
  \end{itemize}
  Now we are ready to prove that the set $\G$ has the required properties. 
  Let $\alpha \in \Gamma^*$ such that 
  $\calP(\alpha, \conf_1 \U \conf_2) \geq \varrho$, and let 
  $\beta = \restrict{\alpha}{S}$. There are two possibilities:
  \begin{itemize}
  \item $\mathit{length}(\beta) < n$. Then 
    $\calP^u(\beta,\conf_1 \U \conf_2) \geq \varrho$, hence $\beta \in \G$
    and $\alpha \in \bigcup_{\beta \in \G} \Gen_n(\beta)$.
  \item $\mathit{length}(\beta) \geq n$. Let $\beta = \gamma\gamma'$ where
    $\mathit{length}(\gamma) = n$. Due to the observation (a) above we have
    that $\calP(\gamma,\conf_1 \U \conf_2) \geq \varrho - \lambda/3$,
    hence also $\calP^u(\gamma,\conf_1 \U \conf_2) \geq \varrho - \lambda/3$,
    which means that
    $\gamma \in \G$ and thus $\alpha \in \bigcup_{\beta \in \G} \Gen_n(\beta)$.
  \end{itemize}
  Now let $\alpha \in \Gen_n(\beta)$ for some $\beta \in \G$. Again, we 
  distinguish two possibilities:
  \begin{itemize}
  \item $\mathit{length}(\beta) < n$. Then 
     $\calP^u(\beta,\conf_1 \U \conf_2) \geq \varrho$, which means that
     $\calP(\beta,\conf_1 \U \conf_2) \geq \varrho - \lambda/3$ by the
     observation (b) above. Hence, 
     $\calP(\alpha,\conf_1 \U \conf_2) \geq \varrho - \lambda/3$.
  \item $\mathit{length}(\beta) = n$. Then 
     $\calP^u(\beta,\conf_1 \U \conf_2) \geq \varrho - \lambda/3$, 
     which means that 
     $\calP(\beta,\conf_1 \U \conf_2) \geq \varrho - 2\lambda/3$
     due to the observation (b). Further, for every $\alpha' \in \Gamma$
     we have that 
     $\calP(\beta\alpha',\conf_1 \U \conf_2) \geq \varrho - \lambda$
     due to the observation~(a) above. Hence, 
     $\calP(\alpha,\conf_1 \U \conf_2) \geq \varrho - \lambda$ as required.
  \end{itemize}

  The automaton $\A^\leq$ is constructed similarly. Here, the
  set $\G$ is computed using the lower approximations $[X,\bullet]^\ell$
  and $[X,\varepsilon]^\ell$. Since this construction is analogous to the
  one just presented, it is not given explicitly.
\end{proof}

\begin{thm}
  There is an error-tolerant PCTL model checking algorithm for pBPA pro\-cesses.
\end{thm}
\begin{proof}
  The proof is similar to the one of Theorem~\ref{thm-PDA-CTL}, using 
  Lemma~\ref{lem-BPA-X} and~\ref{lem-BPA-U} instead of 
  Lemma~\ref{lem-qualit-X},~\ref{lem-qualit-U1},~and~\ref{lem-qualit-U0}.
  Note that Lemma~\ref{lem-reg-sim} is applicable also to pBPA (the system 
  $\Delta'$ constructed in Lemma~\ref{lem-reg-sim} has the same set of 
  control states as the original system $\Delta$).
\end{proof}

\section{Model Checking $\omega$-regular Specifications}
\label{sec-Buchi}

In this section we show that the qualitative and quantitative 
model-checking problem for pPDA and $\omega$-regular properties
are decidable. At the very core of our result are observations 
leading to the definition of a finite Markov chain 
$\MH{\Delta}$. Intuitively, each transition of $\MH{\Delta}$
corresponds to a sequence of transitions of the probabilistic transition system
$\T_\Delta$ associated to $\Delta$. This allows to reduce the model-checking problem
to a problem about $\MH{\Delta}$, which, since $\MH{\Delta}$ is finite,
 can be solved using well-known techniques. 
In \cite{EKM:prob-PDA-PCTL}, the Markov chain $\MH{\Delta}$ was used 
to show that the qualitative and quantitative model-checking problem 
for properties expressible by deterministic B\"{u}chi automata is 
decidable. Later, it was observed in \cite{BKS:pPDA-temporal} that 
the technique can easily be generalized to deterministic Muller
automata. Thus, the decidability result was extended to all $\omega$-regular
properties. In this paper we go a bit further, and prove the decidability
of a slightly larger class. The previous result about the $\omega$-regular
case follows as a corollary.

The section is structured as follows. Given a pPDA $\Delta$, we first
introduce the notion of \emph{minima of a run}
and \emph{$\Delta$-observing automaton}. We use observing automata as
specifications: an infinite run satisfies the specification if{}f it is 
accepted by the automaton (section \ref{sub:min}). Using the notion of minima, 
we define the finite Markov chain $\MH{\Delta}$ (section \ref{sub:chain}), 
and show that the probability that a run is accepted by a
$\Delta$-observing automaton is effectively expressible in $(\mathbb{R},+,*,\leq)$ (section \ref{sub:exp}).
Finally, we show that the model-checking problem for $\omega$-regular 
properties is a special case of the problem of deciding
if a run is accepted by a
$\Delta$-observing automaton with at least a given probability (section \ref{sub:reg}).

For the rest of this section, we fix a pPDA  
$\Delta = (Q,\Gamma,\delta,\Prob)$. 

\subsection{Minima of a run}
\quad \label{sub:min}
Loosely speaking, a configuration of a run 
is a minimum if all configurations placed after it in the run have the same 
or larger stack length.

\begin{defi}
  Let $w = p_1\alpha_1;p_2\alpha_2,\cdots$ be an infinite run in $\T_\Delta$.
  A configuration $p_i\alpha_i$ is a {\em minimum} of $w$ if 
  $|\alpha_i| \leq |\alpha_j|$ for every $j \geq i$. We say that
  $p_i\alpha_i$ is the \emph{$k^{{\it th}}$ minimum} of $w$ if $p_i\alpha_i$  is a minimum
  and there are exactly $k-1$ indices $j < i$ such that $p_j\alpha_j$ is a minimum.
  We denote the $k^{{\it th}}$ minimum of $w$ by $\min_k(w)$.
\end{defi}

\noindent Sometimes we abuse language and use $\min_i(w)$ to denote 
not only a configuration, but the particular \emph{occurrence} of the 
configuration that corresponds to the $i^\mathit{th}$ minimum. 

\begin{exa}
In the run $w_1 =(Z;DZ)^\omega$ of the pBPA shown in the introduction
we have $\min_i(w_1) = Z$ for every $i \geq 1$. In the run
$w_2 = Z;DZ;DDZ;\ldots$ we have $\min_1(w_2) = Z$ and $\min_i(w_2) = D$
for every $i \geq 2$. Every odd configuration of $w_1$ is a minimum,
and every configuration of $w_2$ is a minimum. \hfill$\Box$
\end{exa}

Since stack lengths are bounded from below, every infinite run
has infinitely many minima, and so it can be divided into an infinite
sequence of fragments, or ``jumps'', each of them leading 
from one minimum to the next.

We are interested in those properties of a run 
that can be decided by extracting a finite amount of information from each jump, 
independently of its length. Consider for instance the property
``the control state $p$ is visited infinitely often along the run''.
It can be reformulated as ``there are infinitely many 
jumps along which the state $p$ is visited''. In order to decide the property
all we need is a bit of information for each jump, telling 
whether it is ``visiting'' or ``non-visiting''. We consider properties
in which this finite amount of information can be extracted by letting 
a finite automaton go over the jump reading the {\em heads} of the 
configurations:

\begin{defi}
  Given a configuration $pX\alpha$ of $\Delta$, we call $pX$ the
  \emph{head} and $\alpha$ the \emph{tail} of $pX\alpha$.  The set 
  $Q \times \Gamma$ of all heads of $\Delta$ is also denoted by
  $\H(\Delta)$.
\end{defi}

More precisely, we consider automata with the set of heads as alphabet. 
An oracle tells the automaton
to start reading heads immediately after the run leaves a minimum (i.e., the first head
read is the one of the configuration immediately following the minimum), stop
after reading the head of the next minimum, report its state, and reset itself to an 
initial state that depends on the head of the minimum. 

\begin{defi}
\label{def-observing-automaton}
  A \emph{$\Delta$-observing automaton} is a tuple $\A = (A,\xi, a_o, \Acc)$ where $A$ is
  finite set of \emph{observing states}, 
  $\xi : A \times \H(\Delta) \rightarrow A$ 
  is a (total) \emph{transition function}, $a_0 \in A$ is 
  an \emph{initial state}, and $\Acc$ is a  
  set of subsets of $A$, also called an \emph{acceptance set}.

Let $w$ be an infinite run in $\T_\Delta$ 
and let $i \in \Nset$. The \emph{$i^{\mathit{th}}$ observation}
of $\A$ over $w$, denoted $\Obs_i(w)$, is  the state reached by $\A$ after reading the heads of all  configurations between $\min_{i}(w)$ and 
$\min_{i+1}(w)$, including $\min_{i+1}(w)$ but not including $\min_{i}(w)$.
\footnote{Notice that the automaton starts observing after the first 
minimum of the run.}
The {\em observation} of $\A$ on $w$, denoted by  
$\Obs(w)$, is the sequence $\Obs_1(w)\Obs_2(w) \ldots$. 

We say that an infinite run $w \in \run(pX)$ is \emph{accepting} 
if the set of states of $A$ that occur infinitely 
often in $\Obs(w)$ belongs to ${\it Acc}$; otherwise, $w$ is \emph{rejecting}. 
\end{defi}

\begin{exa}
Figure \ref{fig:obs} shows a $\Delta$-observing automaton for the pBPA 
of the introduction (see also Figure \ref{fig-walk}).
For every infinite run $w$ and every $i \geq 0$, we have $\Obs_i(w) = b$ if 
some configuration
of the $i^{{\it th}}$ jump has $Z$ as topmost stack symbol.
So a run is accepting if{}f it visits configurations with head $Z$ 
infinitely often. \hfill$\Box$
\end{exa}

\begin{figure}
\[\xy;<20 pt,0 pt>:
  (0,0)*{\phantom{\hbox{\em Acc}=\{\{a_1\}\}}},
  (3,0)*+{a_0}*{\cir<8 pt>{}}="A",
  (3,0)*+{a_0}*i{\cir<6.5 pt>{}}="a",
  (6,0)*+{a_1}*{\cir<8 pt>{}}="B",
  (6,0)*+{a_1}*i{\cir<6.5 pt>{}}="b",
  (9,0)*{\hbox{\em Acc}=\{\{a_1\}\}},
  \ar(1.5,0);"A"
  \ar^Z"A";"B"
  \ar@(ur,ul)_{I,D}"a"!UR;"a"!UL
  \ar@(ur,ul)_{Z,I,D}"b"!UR;"b"!UL
\endxy
\]
\caption{An observing automaton} \label{fig:obs}
\end{figure}

For the rest of the section we fix a $\Delta$-observing automaton $\A = (A,\xi,a_0, \Acc)$.
Let $\run(pX,\Acc)$ be the set of all
accepting runs initiated in $pX$. Our aim is to show that
$\calP(\run(pX,\Acc))$ is effectively definable in $(\mathbb{R},+,*,\leq)$.

\subsection{The Markov chain $\MH{\Delta}$.}
\label{sub:chain}
\quad For all $pX \in \H(\Delta)$ and all $i \in \Nset$ we define a random
variable $\rv{i}{pX}$ over $\run(pX)$. Loosely speaking, $\rv{i}{pX}$
assigns to a run starting at the configuration $pX$ the head of its
$i^{th}$ minimum, and the $i^{th}$ observation of the $\Delta$-observing automaton
$\A$. Formally, the possible values
of $\rv{i}{pX}$ are pairs of the form $(qY,a)$, where $qY \in
\H(\Delta)$ and $a \in A$. There is also a special value $\bot$, where
$\bot \not\in \H(\Delta) \times A$. For a given $w \in \run(pX)$, the value 
$\rv{i}{pX}(w)$ is determined as follows: If $w$ is finite, then
$\rv{i}{pX}(w) = \bot$; otherwise, $\rv{i}{pX}(w) = (qY,\Obs_i(w))$,
where $qY$ is the head of $\min_i(w)$. Notice that the random
variables are well defined, because they assign to each run exactly
one value.

Given possible values 
$v_1, \ldots, v_n$ for the variables $\rv{1}{pX}, \ldots, \rv{n}{pX}$,
we are going to prove the following two results:
\begin{itemize}
\item the probability that a run satisfies 
$\rv{1}{pX}=v_1, \ldots, \rv{n}{pX}=v_n$ is expressible in $(\mathbb{R},+,*,\leq)$ 
(Lemma \ref{prob-exists}); and
\item the probability that $\rv{i+1}{pX}=v_{i+1}$ depends only on the
value of $\rv{i}{pX}$, but neither on $i$ nor on the value of $\rv{k}{pX}$ for $k < i$ 
(Lemma \ref{cond} and \ref{cond2}).
\end{itemize}
The second result will allow us to define the finite Markov chain $\MH{\Delta}$, while
the first one will show that its transition probabilities are expressible in $(\mathbb{R},+,*,\leq)$.

The proof of Lemma \ref{prob-exists} is rather technical (as we shall, see,
Lemma \ref{cond} and \ref{cond2} are easy corollaries of Lemma \ref{prob-exists}).
We need three auxiliary lemmas.
Intuitively, the first one states that 
the probability of executing an infinite run from a configuration 
$pX$ is equal to the probability of executing an infinite run from
$pX\beta$ \emph{such that the stack content never goes ``below'' $\beta$}.
For every finite or infinite path $w = p_1\alpha_1; p_2 \alpha_2; \cdots$ 
in $\T_\Delta$ and every $\beta \in \Gamma^*$, the symbol 
$w^{+\beta}$ denotes the path $p_1\alpha_1\beta; p_2\alpha_2\beta; \cdots$
obtained from $w$ by concatenating $\beta$ to the stack content in every
configuration. Similarly, if $R$ is a set of paths in $\T_\Delta$ and
$\beta \in \Gamma^*$, then $[R]^{+\beta}$ denotes the set 
$\{w^{+\beta} \mid w \in R\}$. 

\begin{lem}
\label{lem-concat}
  Let $pX \in Q{\times}\Gamma$ and $\beta \in \Gamma^*$. Then 
  $\calP([\irun(pX)]^{+\beta}) = \calP(\irun(pX))$.
\end{lem}
\begin{proof}
  Let $\textit{Dead} = Q {\times} \{\varepsilon\} \cup \{qY\alpha \mid
  qY \text{ has no transitions in } \delta, \alpha \in \Gamma^*\}$.
  We have that
  \begin{eqnarray*}
  \calP([\irun(pX)]^{+\beta}) & = &
 1-\calP(pX\beta,\conf(\Delta)^\bullet\beta\,\U\,\textit{Dead}\,\beta)\\
  & = & 1 - \calP(pX,\conf(\Delta) \,\U\,\textit{Dead}) 
  \mbox{\hspace*{3ex}(by Lemma~\ref{lem-tail})}\\
  & = & \calP(\irun(pX)).
 \end{eqnarray*}
\end{proof}

The second lemma states that prefixing a measurable set of runs with
a finite path yields a measurable set of runs, and relates the 
probabilities of both sets.

\begin{lem}
\label{lem-odot}
  Let $s_0; \cdots; s_n$ be a path in a probabilistic transition system,
  and let $R$ be a measurable subset of $\run(s_n)$. Then 
  $\{s_0; \cdots; s_n\} \odot R$ is a measurable subset of $\run(s_0)$, and
  moreover $\calP(\{s_0; \cdots; s_n\} \odot R) = 
  \Pi_{i=1}^n x_i \cdot \calP(R)$, where $s_i \ltran{x_{i+1}} s_{i+1}$
  for every $0 \leq i < n$. (The `$\odot$' operator has been introduced
  in Definition~\ref{def-notation}.)
\end{lem}
\begin{proof}
  Standard.
\end{proof}

The third lemma shows that the probability of starting from 
the configuration $qY$ reaching the configuration $q\varepsilon$ with the 
observing automaton in state $a$ is expressible in $(\mathbb{R},+,*,\leq)$.

\begin{defi}
  Let $R$ be a $\calP$-measurable set of runs of $\T_\Delta$ starting at the 
  same initial configuration.
  We say that $\calP(R)$ is \emph{well-definable} if there effectively
  exist a pPDA $\Delta'$ and a finite family of probabilities of the form
  $\calP(\run(qY,\conf_1 \,\U\, \conf_2))$, where $qY \in \H(\Delta')$ and
  $\conf_1,\conf_2 \subseteq \conf(\Delta')$ are simple sets, such that
  $\calP(R)$ is effectively definable from this family of probabilities
  using only summation, multiplication, and rational constants.
\end{defi}
\noindent Note that if $\calP(R)$ is well-definable, it can be expressed in
$(\mathbb{R},+,*,\leq)$ using the results of Section~\ref{sec-random}.

For all $qY \in \H(\Delta)$, $r \in Q$, $Z \in \Gamma$, and $a \in A$, let
$\run(qY,r,Z,a) \subseteq \run(qY)$ be the set of all runs
$w = s_0; \cdots; s_n$ such that $s_0 = qY$, $s_n = r \varepsilon$, 
and the automaton $\A$ reaches the state $a$
after reading the heads of configurations $s_0,\cdots,s_{n-1},rZ$.

\begin{lem}
\label{lem-obs-well}
  $\calP(\run(qY,r,Z,a))$ is well-definable.  
\end{lem}
\begin{proof}
  We put $\Delta' = (Q {\times} A, \Gamma, \delta',\Prob')$ to be the 
  synchronized product of $\Delta$ and $\A$, i.e.,
  $(p,\bar{a})X \tran{x} (t,\hat{a})\alpha$ is a rule of $\Delta'$ iff
  $pX \tran{x} t\alpha$ is a rule of $\Delta$ and 
  $\xi(\bar{a},pX) = \hat{a}$. Let 
  $\bar{A} = \{\bar{a} \in A \mid \xi(\bar{a},rZ) = a\}$.
  Now we can easily check that $\calP(\run(qY,r,Z,a))$ is equal to
  \[
    \calP(\,(q,a_0)Y,\, \conf(\Delta')\, \U \, \{(r,\bar{a})\varepsilon\}) 
  \] 
\end{proof}

We can now prove our main technical result:

\begin{lem}
\label{prob-exists}
  For all $pX \in \H(\Delta)$, $n \in \Nset$, and 
  $v_1,\cdots,v_n \in (\H(\Delta){\times} A) \cup \{\bot\}$, the
  probability of 
  $\rv{1}{pX} {=} v_1 \wedge \cdots \wedge \rv{n}{pX} {=} v_n$
  is well-definable. In particular, for every rational constant $y$ 
  there is an effectively constructible formula of $(\mathbb{R},+,*,\leq)$ 
  which holds if and only if 
  $\calP(\rv{1}{pX} {=} v_1 \wedge \cdots \wedge \rv{n}{pX} {=} v_n) = y$.
\end{lem}
\begin{proof}
  By induction on $n$ we prove that 
  $\calP(\rv{1}{pX} {=} v_1 \wedge \cdots \wedge \rv{n}{pX} {=} v_n)$ is 
  well-definable. The base case when $n=1$ follows immediately,
  because $\calP(\rv{1}{pX} {=} v_1)$ equals either $\calP(\irun(pX))$,
  $1 - \calP(\irun(pX))$, or $0$, depending on whether $v_1 = (pX,a_0)$,
  $v_1 = \bot$, or $(pX,a_0) \neq v_1 \neq \bot$, respectively. 
  Observe that $\calP(\irun(pX)) = 
  1 - \calP(pX, \conf(\Delta) \,\U\, \mathit{Dead})$, where
  $\mathit{Dead} = Q {\times} \{\varepsilon\} \cup \{qY\alpha \mid
  qY \text{ has no transitions in } \delta, \alpha \in \Gamma^*\}$.  

  Now let $n \geq 2$. For each $1 \leq i \leq n$, let
  $\sat_i$ be the set of all runs that satisfy
  $\rv{1}{pX} {=} v_1 \wedge \cdots \wedge \rv{i}{pX} {=} v_{i}$.
  If $\calP(\sat_{n-1}) = 0$, which is decidable by induction hypothesis, 
  then $\calP(\sat_n) = 0$ as well. If $\calP(\sat_{n-1}) \neq 0$ and
  there is an $i \leq n-1$ such that $v_i = {\bot}$, then for all
  $j \leq n-1$ we have that $v_j = {\bot}$, and $\calP(\sat_n)$ is
  equal either to $\calP(\sat_{n-1})$ or $0$, depending on whether
  $v_n = {\bot}$ or not, respectively. If $\calP(\sat_{n-1}) \neq 0$, 
  $v_i \neq {\bot}$ for all $i \leq n-1$, and $v_n = {\bot}$, then
  $\calP(\sat_n) = 0$. So, the only interesting case is
  when $\calP(\sat_{n-1}) \neq 0$ and $v_i \neq {\bot}$ for all $i\leq n$. 
  Since
  \begin{equation*}
     \calP(\sat_n)  = 
     \frac{\calP(\rv{n}{pX} {=} v_n \mid \sat_{n-1})}{\calP(\sat_{n-1})}
  \end{equation*} 
  and $\calP(\sat_{n-1})$ is well-definable by induction hypothesis, it 
  suffices to show that the conditional probability 
  $\calP(\rv{n}{pX} {=} v_n \mid \sat_{n-1})$ is also well-definable.
  For this we use a general result of basic probability 
  theory saying that if $A,B$ are events and $B = \uplus_{i \in I} B_i$,
  where $I$ is a finite or countably infinite index set, then
  \begin{equation*}
     \calP(A \mid B)  = 
     \frac{\sum_{i\in I}\calP(A \mid B_i) \cdot \calP(B_i)}{\calP(B)}
  \end{equation*} 
  An immediate consequence of this equation is that if the probability
  $\calP(A | B_i)$ is independent of $i$, then 
  $\calP(A | B) = \calP(A | B_i)$. In our case, $A$ is 
  the event $\rv{n}{pX} {=} v_n$, and $B$ is $\sat_{n-1}$.
  Let 
  \[
     \chop = \{w(0); \cdots; w(\mathrm{min}_{n-1}(w)) \mid w \in \sat_{n-1}\}.
  \]
  Observe that if $y \in \chop$, then the last configuration of $y$ is
  of the form $p_{n-1}X_{n-1}\alpha$. We denote the $\alpha$ by
  $\Stack(y)$. For every $y \in \chop$, let
  \begin{equation}
    \sat_{n-1}(y)  = \{y\} \odot [\irun(p_{n-1}X_{n-1})]^{+\Stack(y)}
  \label{eq-satn} 
  \end{equation}
  Now we can easily check that
  \begin{equation*}
     \sat_{n-1}  =  \biguplus_{y \in \chop} \sat_{n-1}(y)
  \end{equation*}
  Hence, $\chop$ plays the role of $I$, and $\sat_{n-1}(y)$ 
  plays the role of $B_i$. We show that 
  $\calP(\rv{n}{pX} {=} v_n \mid \sat_{n-1}(y))$ is independent of $y$,
  which means that 
  \begin{equation*}
     \calP(\rv{n}{pX} {=} v_n \mid \sat_{n-1}(y)) = 
     \calP(\rv{n}{pX} {=} v_n \mid \sat_{n-1}).
  \end{equation*}
  By definition of conditional probability, 
  \begin{equation}
    \calP(\rv{n}{pX} {=} v_n \mid \sat_{n-1}(y)) = 
    \frac{\calP(\rv{n}{pX} {=} v_n \wedge \sat_{n-1}(y))}{\calP(\sat_{n-1}(y))}
  \label{eq-goal}
  \end{equation}
  The  denominator of the fraction in equation~(\ref{eq-goal}) 
  is well-definable, because
  \begin{equation*}
    \calP(\sat_{n-1}(y))  = \calP(\run(y)) \cdot \calP(\irun(p_{n-1}X_{n-1}))
  \end{equation*}
  Here we used Lemma~\ref{lem-concat}, Lemma~\ref{lem-odot}, and 
  equation~(\ref{eq-satn}). Now we show that 
  $\calP(\rv{n}{pX} {=} v_n \wedge \sat_{n-1}(y))$ is also well-definable. 
  Let $R$ be the set of
  all runs satisfying $\rv{n}{pX} {=} v_n \wedge \sat_{n-1}(y)$, and let
  $v_n = (p_n X_n,a_n)$ and $\alpha = \Stack(y)$. Obviously, each 
  $w \in R$ starts with $y$. Now
  let us consider what transitions can be performed from the final state
  $p_{n-1}X_{n-1}\alpha$ of $y$. 
  \begin{itemize}
    \item Obviously, transitions which decrease 
        the stack cannot be performed, because 
        $p_{n-1}X_{n-1}\alpha$ would
        not be a minimum then (i.e., $w$ would not belong to $R$). 
    \item If a transition of the form 
        $p_{n-1}X_{n-1}\alpha \tran{x} rZ\alpha$ is performed, then 
        $rZ\alpha$ must be the $n$-th minimum, because the stack cannot be 
        decreased below $Z$ (otherwise, $p_{n-1}X_{n-1}\alpha$ would
        not be a minimum). So, if $w \in R$, we must have that 
        $rZ = p_nX_n$ and $\xi(a_0,rZ) = a_n$.
    \item  If a transition of the form 
        $p_{n-1}X_{n-1}\alpha \tran{x} rPQ\alpha$ is performed, 
        then the stack cannot be decreased below $Q$. Now there
        are two possibilities:
        \begin{itemize}
        \item If the stack is never decreased  below $P$, then 
           the configuration $rPQ\alpha$ is the $n$-th minimum. Hence, if 
           $w \in R$, we must have that  $rP = p_nX_n$ and 
           $\xi(a_0,rP) = a_n$.
        \item If the stack is decreased below $P$, i.e., if a sequence
           of transitions is performed of the form \mbox{$rPQ\alpha \tran{}^* tQ\alpha$}
           (where the stack is never decreased to $Q\alpha$ 
           except in the last configuration), then
           $tQ\alpha$ is the $n$-th minimum. Hence, if $w \in R$, 
           we must have
           that $tQ = p_n X_n$ and the automaton $\A$ reaches $a_n$ 
           by reading 
           the word consisting of heads of configurations in the sequence 
           $rPQ\alpha \tran{}^* tQ\alpha$.
        \end{itemize}
  \end{itemize}
  From the above discussion, it follows that $R$ can be partitioned 
  as follows:
  \begin{eqnarray*}
    R & = & \biguplus_{p_{n-1} X_{n-1} \tran{x} p_n X_n} 
     \{y\} \odot \{p_{n-1}X_{n-1}\alpha;, p_n X_n \alpha\} \odot
     [\irun(p_nX_n)]^{+\alpha}\\ 
   && \biguplus_{\substack{p_{n-1} X_{n-1} \tran{x} p_n X_n Y\\Y \in \Gamma}}
     \{y\} \odot \{p_{n-1}X_{n-1}\alpha;  p_n X_n Y \alpha\} \odot
     [\irun(p_nX_n)]^{+Y\alpha}\\ 
   && \biguplus_{\substack{p_{n-1} X_{n-1} \tran{x} q Y X_n\\q \in Q, 
     Y \in \Gamma}}
     \{y\} \odot [\run(qY,p_n,X_n,a_n)]^{+\alpha} 
     \odot [\irun(p_nX_n)]^{+\alpha}
  \end{eqnarray*}

  \noindent
  Using Lemma~\ref{lem-concat}, Lemma~\ref{lem-odot}, Lemma~\ref{lem-obs-well}
  and the above equation, we obtain that
  \begin{equation*}
    \calP(\rv{n}{pX} {=} v_n \wedge \sat_{n-1}(y))  =  
    \calP(\run(y)) \cdot \calP(\irun(p_nX_n)) \cdot S
  \end{equation*}
  where
  \begin{eqnarray}
   S & = & \sum_{p_{n-1}X_{n-1} \tran{x} p_nX_n} x \quad + \quad
           \sum_{\substack{p_{n-1}X_{n-1} \tran{x} p_nX_nY\\Y \in \Gamma}} 
            x \quad +\nonumber\\
    &&
        \sum_{\substack{p_{n-1}X_{n-1} \tran{x} qYX_n\\q \in Q, Y \in \Gamma}} 
        x \cdot \calP(\run(q Y, p_n,X_n,a_n))
  \label{eq-S}
  \end{eqnarray}
  Equation~(\ref{eq-goal}) can now be rewritten to 
  \begin{equation}
    \calP(\rv{n}{pX} {=} v_n \mid \sat_{n-1}(y))  =  
    \frac{\calP(\irun(p_nX_n))}{\calP(\irun(p_{n-1}X_{n-1}))} \cdot S
  \label{eq-final}
  \end{equation}
  where the meaning of $S$ is given by equation~(\ref{eq-S}). So, 
  $\calP(\rv{n}{pX} {=} v_n \mid \sat_{n-1}(y))$ is indeed independent
  of $y$, and hence equation~(\ref{eq-final}) also defines the probability
  $\calP(\rv{n}{pX} {=} v_n \mid \sat_{n-1})$.
\end{proof}

\noindent
Loosely speaking, the following lemma proves the memoryless
property required to define a Markov chain: The probability
of $\rv{n}{pX} = v_n$ depends only on the value of
$\rv{n-1}{pX}$, not on the values of $\rv{n-2}{pX}, \ldots, \rv{1}{pX}$.

\begin{lem}
\label{cond}
The conditional probability of $\rv{n}{pX} = v_n$ on the hypothesis
$\rv{1}{pX} = v_1 \wedge \cdots \wedge \rv{n-1}{pX} = v_{n-1}$ is
equal to the probability of $\rv{n}{pX} = v_n$ conditioned on
$\rv{n-1}{pX} = v_{n-1}$, assuming that the probability of $\rv{1}{pX}
= v_1 \wedge \cdots \wedge \rv{n-1}{pX} = v_{n-1}$ is non-zero.
\end{lem}
\begin{proof}
The result follows
immediately from Equation~(\ref{eq-final}) in the proof of 
Lemma~\ref{prob-exists}: The right side on the equation does not depend
on the values of $\rv{n-2}{pX}, \ldots, \rv{1}{pX}$.
\end{proof}

Finally, as another consequence of 
Lemma~\ref{prob-exists} we obtain that the probability
of $\rv{n}{pX} = v_n$ does not depend on $n$:

\begin{lem}
\label{cond2}
  The conditional probability of $\rv{n}{pX} = (q'Y',a')$ on the
  hypothesis $\rv{n-1}{pX} = (qY,a)$ is equal to the conditional
  probability of $\rv{2}{qY} = (q'Y',a')$ on the hypothesis
  $\rv{1}{qY} = (qY,a_0)$, assuming that $\calP(\rv{n-1}{pX} =
  (qY,a)) \neq 0$. Moreover, the hypothesis that a run $w$
  satisfies $\rv{1}{qY}(w) = (qY,a_0)$
  is the same as the hypothesis that $w \in \irun(qY)$.
\end{lem}
\begin{proof}
The first part follows immediately from the fact that
$n$ appears only as an index in Equation~(\ref{eq-final}).
For the second, observe that,
by definition, a run $w$ starting at $qY$
satisfies $\rv{1}{qY}(w) {=} qY$ if (1) it is infinite and (2) 
its first minimum has head $qY$. But (1) and the fact that 
all configurations of an infinite run have length
1 or greater imply that the first configuration of the run is also its
first minimum, and so, since $w$ starts at $qY$, they imply (2). So a 
run $w$ starting at $qY$ satisfies $\rv{1}{qY} {=} qY$ if{}f it is infinite,
i.e., if{}f $w \in \irun(qY)$.
\end{proof}

\begin{exa}
In order to give some intuition for these results, and in particular for 
the proof of Lemma \ref{prob-exists}, consider the special
case in which the initial configuration is $pX$ for some
$p\in P, X \in \Gamma$, and the observing automaton $\A$ has one single 
state. In this case, the automaton always makes the same observation,
and so  we can write  $\rv{n}{pX} = qY$ instead of 
$\rv{n}{pX} = (qY, a)$. We wish to obtain an expression for 
$\calP(\rv{2}{pX} {=} qY)$. By the second part of Lemma \ref{cond2}
we have
\[ \calP(\rv{1}{pX} {=} pX) = \calP(\irun(pX)) \]
and therefore
\[ \calP(\rv{2}{pX} {=} qY) = \calP(\rv{2}{pX} {=} qY \mid \rv{1}{pX} {=} pX) \cdot \calP(\irun(pX)) \]
\noindent Now we can apply equation \ref{eq-S} in the proof
of Lemma \ref{prob-exists} and obtain

\[\calP(\rv{2}{pX} {=} qY) = \calP(\irun(qY)) \cdot S \]

\noindent and, by Equation \ref{eq-goal} 

\begin{eqnarray}
\calP(\rv{2}{pX} {=} qY) & = &
\sum_{pX \tran{x} qY} x  \cdot \calP(\irun(qY)) \quad + \quad \nonumber \\
& & \sum_{\substack{pX \tran{x} qYZ\\Z \in \Gamma}} 
            x \cdot \calP(\irun(qY)) \quad + \quad \nonumber \\
& & \sum_{\substack{pX \tran{x} rZY \\r \in Q, Z \in \Gamma}} 
        x \cdot \calP(rZ, (Q \times \Gamma^*)\U \{q\varepsilon\}) \cdot \calP(\irun(qY)) 
\label{eq-ex}
\end{eqnarray}

Let us interpret this equation. In order to reach the second minimum at 
$qY$ there are only three possibilities for the first move. 
The first possibility is to move directly from $pX$ to $qY$; in this case 
we must
continue with any run that never terminates, since every infinite run 
of the form $pX;qY;\cdots$ necessarily has $qY$ as second minimum. The probability
of this case is captured by the first summand of Equation \ref{eq-ex}. 
The second possibility is to move from $pX$ to $qYZ$ 
for some $Z \in \Gamma$; in this case 
we must continue with an infinite run in which the stack content 
always has at least length 2,
i.e., with a run of the form 
\[pX;\, qYZ;\, q_1\alpha_1Z; \, \ldots \, ; \, q_i\alpha_iZ;\,  \ldots\]
\noindent where all the $\alpha$'s are nonempty.
This gives the second summand. Finally, the third possibility is to move 
from $pX$ to $rZY$ for some $r\in P, Z \in \Gamma$; we must then continue 
with a run that eventually ``pops the $Z$'' while entering state $q$, i.e.,
with a run of the form 
\[ pX;\, rZY;\, r_1\alpha_1Y;\,  \ldots \,;\, r_n\alpha_nY;\, qY;\, q_1\beta_1;\, \ldots; \, q_i\beta_i ; \, \ldots \]
\noindent where all the $\alpha$'s and $\beta$'s are nonempty. This gives the third summand. \hfill$\Box$
\end{exa}

Lemma \ref{cond} and \ref{cond2} allow us to define the finite Markov chain $\MH{\Delta}$.

\begin{defi}
\label{def-MC}
  The finite-state Markov chain $\MH{\Delta}$ has the following 
  set of states
  \[
    \{(qY,a) \mid qY \in \H(\Delta), a \in A,
        \calP(\rv{1}{qY}{=}(qY,a_0)) > 0 \}
    ~~\cup~~ \H(\Delta) ~~\cup~~ \{\bot\}
  \]
  and the following transition probabilities:
  \begin{itemize} 
    \item $\Prob({\bot} \tran{} {\bot}) = 1$, 
    \item $\Prob(pX \tran{} (qY,a_0)) = \calP(\rv{1}{pX} {=} (qY,a_0))$,
    \item $\Prob(pX \tran{} {\bot}) = \calP(\rv{1}{pX} {=} {\bot})$, 
    \item $\Prob((qY,a) \tran{} (q'Y',a')) = 
      \calP(\rv{2}{qY} {=} (q'Y',a') \mid \rv{1}{qY} {=} (qY,a_0))$.
  \end{itemize}
\end{defi}
One can readily check that $\MH{\Delta}$ is indeed a Markov chain, i.e.,
for every state $s$ of $\MH{\Delta}$ we have that the sum of probabilities of
all outgoing transitions of $s$ is equal to one. Observe
also that if both $(qY,a)$ and $(qY,a')$ are states of $\MH{\Delta}$, 
then they have the ``same'' outgoing arcs (i.e., 
$(qY,a) \tran{x} (rZ,\bar{a})$ if{}f $(qY,a') \tran{x} (rZ,\bar{a})$, 
where $x > 0$). 

\begin{exa}
We construct the Markov Chain $\MH{\Delta}$ for the
pBPA $\Delta$ of Figure \ref{fig-walk} and the observing automaton 
$\A$ of Figure \ref{fig:obs}. In fact, as we shall see,
the states and transition probabilities
of the chain depend on the value of the parameter $x$.

Since the pBPA has one single control state,
we omit it. The set of heads is then $\H(\Delta) = \{Z,I,D\}$ and
the set of states of the observing automaton is $A = \{a_0, a_1\}$. In order
to determine the states of the Markov chain we have to 
compute the pairs $(Y, a)$ such that $\calP(\rv{1}{Y} = (Y,a)) \geq 0$.
Recall the definition of $\calP(\rv{1}{Y} = (Y,a))$. This is the 
probability of, starting at the configuration $Y$, executing
an infinite run such that ($i$) the head of the first minimum is $Y$, and
($ii$) the first observation of $\A$ is the state $a$.
Since the initial configuration $Y$ has the shortest possible length in 
an infinite run, ($i$) always holds. So $\calP(\rv{1}{Y} = (Y,a))$
is the probability of executing an infinite run such that ($ii$) holds.
Recall that the first observation of an observing automaton
is the state it reaches after
reading the sequence of heads between the first and the second minimum,
excluding the first, but including the second. In the case of
the automaton $\A$ of Figure \ref{fig:obs}, the first
observation is $a_0$ if the sequence of heads does not contain the head $Z$,
and $a_1$ otherwise.

The values of $\calP(\rv{1}{X} = (X, a))$ for $X \in \{Z,I,D\}$ and 
$a \in  \{a_0, a_1\}$ are as follows:

$$
\calP(\rv{1}{X} = (X,a)) = \left\{
\begin{array}{lllll}
\min\{2x, 2 - 2x \} & \mbox{ if} &  X=Z & \mbox{and} & a = a_1\\
\max\{0, (2x - 1)/x\} & \mbox{ if } &  X=I & \mbox{and} &  a = a_0\\
\max\{0, (1 - 2x)/(1-x)\} & \mbox{ if} &  X=D & \mbox{and} &  a = a_0\\
0 & \multicolumn{2}{l}{\mbox{ otherwise}} &
\end{array}
\right.
$$

These values can be obtained using the definitions, but in this simple case 
we can also use more direct methods. Consider  
for instance $\calP(\rv{1}{Z} = (Z,a_1))$. This is the probability of, 
starting at $Z$,
executing an infinite run and visiting again a configuration with head $Z$ 
before reaching the second minimum. Observe that
all runs that start at $Z$ are infinite, that the only 
configuration they visit with head $Z$ is $Z$ itself, and that
$Z$ is always a minimum. 
So $\calP(\rv{1}{Z} = (Z,a_1))$ is the probability 
of, starting at the configuration $Z$, eventually reaching $Z$ again. 
This probability is equal to $x \cdot [I,\varepsilon] + (1-x) \cdot [D, \varepsilon]$, where $[I,\varepsilon]$ and $[D,\varepsilon]$ are defined in 
Example \ref{exa-ber}. We get

$$\begin{array}{rcl}
\calP(\rv{1}{Z}  = (Z,a_1)) & = & x \cdot [I,\varepsilon] + (1-x) \cdot [D, \varepsilon] \\
&= & x \cdot \min\{1, (1-x)/x\} + (1-x) \cdot \min\{1, x/(1-x)\} \\
&= & \min\{2x, 2 - 2x \}
\end{array}
$$

Observe that the states of $\MH{\Delta}$ depend on $x$. The states are
$\bot, Z, I, D$ and
\begin{center}
\begin{tabular}{ll}
$(D, a_0)$ & if $x = 0$,\\
$(Z, a_1), (D, a_0)$ & if $0 < x < 1/2$, \\
$(Z, a_1)$ & if $x = 1/2$, \\
$(Z, a_1), (I, a_0)$ & if $1/2 < x < 1$,\\
$(I, a_0)$ & if $x=1$.
\end{tabular}
\end{center}
\noindent The Markov chain for the cases $x = 1/2$ and  $1/2 < x < 1$ are
shown in Figure \ref{fig:markov}. 
\begin{figure}[ht]
\[\vcenter{\xy;<23 pt,0 pt>:
  (2,2)*+{Z}      ="Z",
  (4,2)*+{(Z,a_1)}="A",
  (4,2)*+i{l}     ="a",
  (0,1)*+{\bot}   ="B",
  (0,1)*+i{-}     ="b",
  (2,1)*+{I}      ="I",
  (2,0)*+{D}      ="D",
  \ar^-1"Z";"A"
  \ar@(ul,ur)^1"a"!UL;"a"!UR
  \ar@(dl,ul)^1"b"!DL;"b"!UL
  \ar_1"I";"B"
  \ar^1"D";"B"  
  \endxy}
\qquad
  \vcenter{\xy;<28 pt,0 pt>:
  (2,3)*+{Z}        ="Z",
  (4,3)*+{(Z,a_1)}  ="A",
  (4,3)*+i{l}       ="a",
  (0,1.5)*+{\bot}   ="B",
  (0,1.5)*+i{-}     ="b",
  (2,1.5)*+{I}      ="I",
  (4,1.5)*+{(Z,a_0)}="C",
  (4,1.5)*+i{l}     ="c",
  (2,0)*+{D}        ="D",
  \ar        ^-{2-2x}          "Z"   ;"A"
  \ar        _(.333){2x-1}     "Z"   ;"C"
  \ar        ^{2x-1}           "A"   ;"C"
  \ar@(ul,ur)^{2-2x}           "a"!UL;"a"!UR
  \ar@(dl,ul)^1                "b"!DL;"b"!UL
  \ar        _{\frac{1-x}{x}}  "I"   ;"B"
  \ar        ^1                "D"   ;"B"  
  \ar        _-{\frac{2x-1}{x}}"I"   ;"C"
  \ar@(dl,dr)_1                "c"!DL;"c"!DR
  \endxy}
\]
\caption{The Markov chain $\MH{\Delta}$ for $x=1/2$ (left)
and for $1/2 < x < 1$ (right) } \label{fig:markov}
\end{figure}
\end{exa}
\noindent Let us obtain the transition probability from $(I,a_0)$ to 
itself in the case $1/2 < x < 1$. According
to Definition \ref{def-MC}, the probability is equal 
to $\calP(\rv{2}{I}=(I,a_0) \mid \rv{1}{I}=(I,a_0))$,
i.e., to the probability of, assuming the first minimum has
head $I$, reaching the
second minimum at head $I$ again, visiting no configuration with head $Z$
in-between. Let us see that this probability is 1. 
If the first minimum is $I\alpha$ for some $\alpha \in \{Z,I,D\}^*$, 
then all
subsequent configurations of the run are of the form $\beta\alpha$ for 
a nonempty $\beta$ (notice that we assume that the run is infinite,
because finite runs have no minima). So $\beta$ must have head $I$
and so, in particular, the next minimum will 
also have head $I$. \hfill$\Box$

Not every run of $\Delta$ is ``represented'' in the Markov chain $\MH{\Delta}$.
Consider for instance the case $x=1/2$ and
its corresponding chain $\MH{\Delta}$ on the left of Figure \ref{fig:markov}.
Every configuration of the run $Z;IZ;IIZ;IIIZ; \ldots$ is a minimum,
but its sequence of heads, i.e., $Z I^\omega$, does not correspond to 
any path of $\MH{\Delta}$. We show, however, that the ``not represented'' 
runs have probability 0.

A \emph{trajectory} in $\MH{\Delta}$ is an infinite sequence
$\sigma(0) \sigma(1) \cdots$ of states of $\MH{\Delta}$, where for
every $i \in \Nset_0$, \mbox{$\Prob(\sigma(i) \tran{} \sigma(i+1)) >
0$}.  To every run $w \in \run(pX)$ of $\Delta$ we associate its
\emph{footprint}, denoted $\sigma_w$, which is an infinite sequence of
states of $\MH{\Delta}$ defined as follows:
\begin{itemize}
\item $\sigma_w(0) = pX$
\item if $w$ is finite, then for every $i \in \Nset$ we have 
  $\sigma_w(i) = {\bot}$;
\item if $w$ is infinite, then for every $i \in \Nset$ we have 
  $\sigma_w(i) = (p_iX_i,\Obs_i(w))$, where
  $p_iX_i$ is the head of $\min_i(w)$.
\end{itemize}
We say that a given $w \in \run(pX)$ is \emph{good} if $\sigma_w$
is a trajectory in $\MH{\Delta}$.
Our next lemma reveals that almost all runs are good.
\begin{lem}
\label{lem-good}
  Let $pX \in \H(\Delta)$, and let \textit{Good} be the subset of all good
  runs of $\run(pX)$. Then $\calP(\textit{Good}) = 1$.
\end{lem}
\begin{proof}
Let $\textit{Bad} = \run(pX) \smallsetminus \textit{Good}$. 
Let $\textit{Fail}$ be the set of all finite sequences 
$v_0 \cdots v_{i+1}$ of states of $\MH{\Delta}$ such that $i \in \Nset_0$,
$v_0 = pX$, $v_0 \cdots v_i$ is a trajectory in $\MH{\Delta}$, and
$\Prob(v_i \tran{} v_{i+1}) = 0$, where  $\Prob$ is the probability 
assignment of $\MH{\Delta}$. Each $y \in \textit{Fail}$ determines
a set $\textit{Bad}_y = \{w \in \textit{Bad} \mid \sigma_w \text{ starts
with } y\}$. Obviously, 
$\textit{Bad} = \biguplus_{y \in \textit{Fail}} \textit{Bad}_y$.
We prove that $\calP(\textit{Bad}_y) = 0$ for each $y \in \textit{Fail}$.
Let $y = v_0 \cdots v_{i+1}$.
By applying definitions, we obtain
{\small\begin{eqnarray*}
\calP(\textit{Bad}_y) & = & \calP(\rv{1}{pX} {=} v_1 \wedge 
\cdots \wedge \rv{i+1}{pX} {=} v_{i{+}1})\\[1ex]
& = &
\frac{\calP(\rv{i+1}{pX} {=} v_{i{+}1} \mid \rv{i}{pX} {=} v_{i} \wedge \cdots
\wedge  \rv{1}{pX} {=} v_{1})}{\calP(\rv{i}{pX} {=} v_{i} \wedge \cdots 
\wedge \rv{1}{pX} {=} v_{1})}
\end{eqnarray*}}
Since $\calP(\rv{i}{pX} {=} v_i \wedge \cdots \wedge \rv{1}{pX} {=} v_{1}) 
\neq 0$, the last fraction makes sense and it is equal to 
\[
  \frac{\Prob(v_i \tran{} v_{i+1})}{\calP(\rv{i}{pX} {=} v_{i} \wedge \cdots 
     \wedge \rv{1}{pX} {=} v_{1})} 
\]
which equals zero.
\end{proof}

\subsection{$\calP(\run(pX,\Acc))$ is effectively definable in $(\mathbb{R},+,*,\leq)$}
\label{sub:exp}
\quad Recall that our aim is to show that
$\calP(\run(pX,\Acc))$ is effectively definable in $(\mathbb{R},+,*,\leq)$.
We will achieve this in Theorem \ref{thm-acc-dec} as an easy corollary
of Lemma \ref{mainprop}. This lemma states that $\calP(pX,\Acc)$ is 
the probability of, starting at $pX$, hitting so-called \emph{accepting} bottom
strongly connected component of $\MH{\Delta}$. As usual,
a strongly connected component of $\MH{\Delta}$ is a maximal set of
mutually reachable states, and bottom strongly connected components are those
from which no other strongly connected components can be reached.

\begin{defi}
Let $C$ be a bottom strongly connected component of $\MH{\Delta}$.
We say that $C$ is \emph{accepting} if $C \neq \{\bot\}$ and the set 
$\{a \in A \mid (qY,a) \in C \mbox{ for some } qY \in \H(\Delta)\}$
is an element of $\Acc$ (remember that $\Acc$ is the acceptance set
introduced after Definition~\ref{def-observing-automaton}). 
Otherwise, $C$ is \emph{rejecting}.

We say that a given pair $(qY,a)$, where  $qY \in \H(\Delta)$ 
and $a \in A$, is \emph{recurrent}, if it belongs to some 
bottom strongly connected component of $\MH{\Delta}$.
\end{defi}

We say that a run $w \in \run(pX)$ \emph{hits} a pair 
$(qY,a) \in \H(\Delta) {\times} A$
if there is some $i \in \Nset$ such that the head of $\min_i(w)$
is $qY$ and $\Obs_i(w) = a$. The next lemma says that an infinite run 
eventually hits a recurrent pair. In this lemma and the next we use
the following well-known results for finite Markov chains (see e.g.
\cite{Feller:book}): 
\begin{itemize}
\item A run visits some bottom strongly connected component of the chain
with probability 1.
\item If a run visits some state of a bottom strongly connected 
component $C$, then it visits all states of $C$ infinitely often
with probability 1.
\end{itemize}

\begin{lem}
\label{lem-hits-rec-one}
  Let us assume that $\calP(\irun(pX)) > 0$. Then the conditional probability 
  that $w \in \run(pX)$ hits a recurrent pair 
  on the hypothesis that $w$ is infinite is equal to one.
\end{lem}
\begin{proof}
  Let $\textit{Rec}$ denote the event that a run of $\run(pX)$ hits
  a recurrent pair.
  Due to Lemma~\ref{lem-good}, we have that
  \begin{equation}
    \calP(\textit{Rec} \mid \irun(pX)) = 
    \calP(\textit{Rec} \mid \irun(pX) \cap \textit{Good})
  \label{eq-hits-rec}
  \end{equation} 
  A run belongs to $\irun(pX) \cap \textit{Good}$ if{}f its footprint
  is a trajectory in $\MH{\Delta}$ that does not hit the state
  $\bot$. A run $w \in \irun(pX) \cap \textit{Good}$ satisfies
  \textit{Rec} if{}f its footprint hits (some) recurrent pair $(qY,a)$.
  It follows directly from the definition
  of $\MH{\Delta}$ that the right-hand side of equation~(\ref{eq-hits-rec}) is
  equal to the probability that a trajectory from $pX$ in $\MH{\Delta}$ 
  hits a bottom strongly connected component on the
  hypothesis that the state $\bot$ is not visited. Since $\MH{\Delta}$ is 
  finite, this happens with probability one. 
\end{proof}

So, an infinite run eventually hits a recurrent pair. Now we prove
that if this pair belongs to an accepting/rejecting bottom strongly connected
component of $\MH{\Delta}$, then the run will be
accepting/rejecting with probability one. 

\begin{lem}
\label{lem-accrej-one}
  The conditional probability 
  that $w \in \run(pX)$ is accepting/rejecting on the hypothesis that the
  first recurrent pair hit by $w$ belongs to an accepting/rejecting 
  bottom strongly connected component of $\MH{\Delta}$ is equal to one.
\end{lem}
\begin{proof}
  The argument is similar as in the proof of
  Lemma~\ref{lem-hits-rec-one}.  Let $C$ be a bottom strongly
  connected component of $\MH{\Delta}$.  By ergodicity, the
  conditional probability that an infinite trajectory in $\MH{\Delta}$
  hits each state of $C$ infinitely often on the hypothesis that the
  trajectory hits $C$ is equal to one.
\end{proof}

\noindent
A simple consequence of Lemma~\ref{lem-accrej-one} is:
\begin{lem}
\label{mainprop}
(cf. Proposition 4.1.5 of \cite{CY:probab-verification-JACM})
Let $pX \in \H(\Delta)$. $\calP(pX,\Acc)$ is equal to the probability
that a trajectory from $pX$ in $\MH{\Delta}$ hits an accepting bottom
strongly connected component of $\MH{\Delta}$.
\end{lem}

\begin{exa}
Consider the pBPA of Figure \ref{fig-walk} and the observing
automaton of Figure \ref{fig:obs}. $\calP(Z,\Acc)$ is the probability
of, starting at $Z$, executing a run that visits configurations with 
head $Z$ infinitely
often. In the case $x = 1/2$, the bottom strongly connected components
of $\MH{\Delta}$ are $\{\bot\}$ and $\{ (Z, a_1) \}$, which are
rejecting and accepting, respectively. Starting at the state $Z$
of $\MH{\Delta}$, the probability
of hitting $\{ (Z, a_1) \}$ is 1, and so $\calP(Z,\Acc)=1$.
In the case $1/2 < x < 1$, the bottom strongly connected components
of $\MH{\Delta}$ are $\{\bot\}$ and $\{ (I, a_0) \}$, which are both 
rejecting, and so $\calP(Z,\Acc) = 0$.
\end{exa}

Since the probability of hitting a given bottom strongly connected component
of a given finite-state Markov chain is effectively definable in
$(\mathbb{R},+,*,\leq)$ by the results of Section \ref{sec-random}, 
and the transition probabilities in $\MH{\Delta}$ are well-definable too, 
we can conclude the following:

\begin{thm}
\label{thm-acc-dec}
  $\calP(\run(pX,\Acc))$ is effectively expressible in 
  $(\mathbb{R},+,*,\leq)$. In particular, for every rational constant
  $y$ and every ${\sim} \in \{{\leq},{<},{\geq},{>},{=}\}$
  there effectively exists a formula of  
  $(\mathbb{R},+,*,\leq)$ which holds if{}f $\calP(\run(pX,\Acc)) \sim y$.
\end{thm} 

\vspace{0.3cm}
\subsection{Decidability of $\omega$-regular properties}
\label{sub:reg}
\quad As a simple corollary of Theorem \ref{thm-acc-dec}, we obtain the 
decidability of the qualitative/quantitative model-checking problem for 
pPDA and $\omega$-regular properties. 
Recall that a language of infinite words over a finite alphabet is 
{\em $\omega$-regular} if{}f it can be accepted by a (deterministic)
Muller automaton.

\begin{defi}
  A deterministic \emph{Muller automaton} is a tuple 
  $\B = (\Sigma,B,\varrho, b_I, \F)$, where $\Sigma$ is a finite 
  \emph{alphabet}, $B$ is a finite set of \emph{states}, 
  $\varrho \colon B \times \Sigma \rightarrow B$ is a
  (total) \emph{transition function} (we write $b \tran{a} b'$ instead of
  $\varrho(b,a) = b'$), $b_I$ is the \emph{initial state}, and 
  $\F \subseteq 2^B$ is a set of \emph{accepting sets}. 

  An infinite word $w$ over the alphabet $\Sigma$ is
  \emph{accepted} by $\B$ if $\mathit{Inf}(w) \in \F$,
  where $\mathit{Inf}(w)$ is the set of all $b \in B$ that appear
  infinitely often in the unique run of $\B$ over the word $w$. 
\end{defi}
We consider specifications given by Muller automata
having $\H(\Delta)$ as their alphabet. It is well known that every LTL
formula whose atomic propositions are interpreted over simple sets can
be encoded into a deterministic Muller automaton having
$\H(\Delta)$ as alphabet. Our results can be
extended to atomic propositions interpreted over arbitrary regular
sets of configurations using the same technique as in
\cite{EKS:PDA-regular-valuations-IC}.

Let us fix a deterministic Muller automaton $\B = (\H(\Delta),B,\varrho, b_I, \F)$.
An infinite run $w$ of $\T_\Delta$ is \emph{accepted by $\B$} if the
associated sequence of heads of configurations in $w$ is accepted 
by $\B$. Let $\run(pX,\B)$ be the set of all $w \in \run(pX)$
that are accepted by $\B$. We show that $\run(pX,\B)$ is effectively
expressible in $(\mathbb{R},+,*,\leq)$, and so we can decide if it is larger than,
smaller than, or equal to some threshold $\rho$.

Loosely speaking, we proceed as follows. We compute the 
synchronized product $\Delta'$ of $\Delta$ and $\B$. Then, 
we define a $\Delta'$-observing automaton $\A$ whose states are sets
of states of $\B$. The automaton observes heads of $\Delta'$, 
which are of the form $(p, b)X$, where
$pX$ is a head of $\Delta$ and $b$ is a state of $b$. At the end of
a ``jump'', $\A$ returns the set of states of $B$ that were visited during
the jump. Hence, the observation $\Obs(w)$ of the automaton on a run $w$ is
a sequence $B_1B_2\ldots$ of sets of states of $B$ containing full information
about which states were visited in which jump. Now it is just a matter
of setting the acceptance set of $\A$ adequately:
The acceptance sets of $\A$ are the sets $\{b_1, \ldots, b_n\}$
of states of $\A$ such that the union $b_1 \cup \ldots \cup b_n$ is an
element of $\F$.

\begin{thm} 
\label{thm-Buchi-dec}
  $\calP(\run(pX,\B))$ is effectively expressible in 
  $(\mathbb{R},+,*,\leq)$. In particular, for every rational constant
  $y$ and every ${\sim} \in \{{\leq},{<},{\geq},{>},{=}\}$
  there effectively exists a formula of  
  $(\mathbb{R},+,*,\leq)$ which holds if{}f $\calP(\run(pX,\B)) \sim y$.
  (Hence, for each $0 < \lambda < 1$ we can compute
  rationals $\calP^\ell,\calP^u$ such that 
  $\calP^\ell \leq  \calP(pX,\Acc)\leq \calP^u$ and 
  $\calP^u - \calP^\ell \leq \lambda$.)
\end{thm}
\begin{proof}
  Let $\Delta' = (Q {\times} B, \Gamma, \delta',\Prob')$ be the 
  synchronized product of $\Delta$ and $\B$, i.e.,
  $(p,b)X \tran{x} (t,b')\alpha$ is a rule of $\Delta'$ iff
  $pX \tran{x} t\alpha$ is a rule of $\Delta$ and 
  $\varrho(b,pX) = b'$. Consider the $\Delta'$-observing 
  automaton $\A = (A,\xi,\I, \Acc)$ where
  $A = 2^B$, $a_0 = \emptyset$, 
  $\xi(M,(p,b)Y) = M \cup \{b\}$ for all $M \subseteq B$ and
  $(p,b)Y \in \H(\Delta')$, and $\Acc$ is defined as follows:
  for every $a_1, \ldots, a_n \in 2^B$, 
  $\{ a_1, \ldots, a_n \} \in \Acc$ if{}f $a_1 \cup \ldots \cup a_n \in \F$.

It is easy to check that
  \[
     \calP(\run(pX,\B)) = \calP(\run((p,b_I)X,\Acc))
  \]
  Now it suffices to apply Theorem~\ref{thm-acc-dec}.
\end{proof}

\section{Conclusions}
\label{sec-concl}
We have provided model checking algorithms for probabilistic pushdown automata
against PCTL specifications, and against $\omega$-regular specifications
represented by Muller automata. Contrary to the case of
probabilistic finite automata, qualitative properties (i.e., whether a property
holds with probability 0 or 1), depend on the exact values of
transition probabilities. 

There are many possibilities for future work. An obvious question is
what is the complexity of the obtained algorithms. 
Of course, this depends on the complexity of the corresponding
fragments of first order arithmetic of reals.
It is known that the fragment obtained by fixing the alternation depth
of quantifiers is decidable in exponential time  
\cite{Grigoriev:Tarski-exponential-JSC}, and that the existential
fragment (and hence also the universal fragment) is decidable even 
in polynomial space 
\cite{Canny:Tarski-exist-PSPACE}. The formulas constructed 
in Section~\ref{sec-random} have a fixed alternation depth, and so
we can conclude that the qualitative/quantitative random walk
problem is decidable in exponential time. Actually, we can do even 
better---if we are interested whether 
$\calP(pX,\conf_1 \U \conf_2) \leq \varrho$, we can simply ask
if there is \emph{some} solution of the corresponding system of 
quadratic equations (cf. Theorem~\ref{thm-least-fix}) such that the component
of the solution which corresponds to $\calP(pX,\conf_1 \U \conf_2)$
is less than or equal to $\varrho$. Obviously, the minimal solution
(i.e., the probability of $\calP(pX,\conf_1 \U \conf_2)$) can only
be smaller. Hence, the existential fragment is sufficient for
deciding whether $\calP(pX,\conf_1 \U \conf_2) \leq \varrho$, and 
similarly we can use the universal fragment to decide
whether $\calP(pX,\conf_1 \U \conf_2) \geq \varrho$. To sum up, the
problem whether $\calP(pX,\conf_1 \U \conf_2) \sim \varrho$, where
${\sim} \in \{{<},{\leq},{>},{\geq},{=}\}$, is decidable in polynomial
space. 

Recently, deeper results concerning the complexity of the reachability
problem for pPDA and pBPA have been presented by Etessami and Yannakakis in 
\cite{EY:RMC-SG-equations}. In particular, they show that 
the \emph{qualitative} 
reachability problem for pBPA processes (i.e., the question whether a
given configuration is visited with probability $1$) is 
decidable in polynomial time.   
It is also shown that the \textsc{Square-Root-Sum} problem 
(i.e., the question whether $\sum_{i=1}^n \sqrt{a_i} \leq c$ for 
a given tuple $(a_1,\ldots,a_n,c)$ of natural numbers)
is polynomially
reducible to the \emph{quantitative} reachability problem for pBPA, and to
the \emph{qualitative} reachability problem for pPDA. The complexity
of the \textsc{Square-Root-Sum} problem is a famous open problem in the
area of exact numerical algorithms. It is known that the problem 
is solvable in polynomial space,
but no lower  bound (like NP or co-NP hardness) is known. This means that the
PSPACE upper bound for the quantitative pBPA reachability and the
qualitative pPDA reachability cannot be improved without achieving
an improvement in the complexity of the \textsc{Square-Root-Sum} problem.

Some of the problems which were left open in \cite{EKM:prob-PDA-PCTL}
were solved later in \cite{BKS:pPDA-temporal}. 
It was shown that the model-checking problems
for PCTL and pPDA, and for PCTL$^*$ and pBPA, are undecidable
(PCTL$^*$ is the probabilistic extension of CTL$^*$). On the 
other hand, the decidability result about qualitative/quantitative 
model-checking pPDA against deterministic B\"{u}chi specifications
was extended to Muller automata. In the qualitative case,
the algorithm runs in time which is singly exponential in the size
of a given pPDA and a given Muller automaton. In the
quantitative case, the algorithm needs exponential space.
Finally, it was shown that the model-checking problem for the
qualitative fragment of the logic PECTL$^*$ and pPDA processes
is also decidable. The complexity bounds are essentially the same
as for Muller properties.

The complexity of model-checking $\omega$-regular properties (encoded
by B\"{u}chi automata) for pPDA and pBPA processes was studied also 
in \cite{EY:pPDA-LTL-complexity}. The complexity bounds
improve the ones given in \cite{BKS:pPDA-temporal}. 
In particular, it is shown that
the qualitative model-checking problem for pPDA and B\"{u}chi 
specifications is EXPTIME-complete. 

An interesting open problem is the decidability of the model-checking
problem for PCTL and pBPA processes, i.e., whether there is an
``exact'' algorithm apart from the error-tolerant one given in
Section~\ref{sec-PCTL-nBPA}. Another area of open problems is
generated by considering model-checking problems for a more general
class of pushdown automata whose underlying semantics is defined in
terms of Markov decision processes (this model combines the paradigms
of non-deterministic and probabilistic choice).
 
\section{Acknowledgments}
The authors would like to thank Stefan Schwoon 
and two anonymous referees for many helpful insights
and comments.

\bibliographystyle{myalpha}
\bibliography{str-long,concur}

\end{document}